Annals of
# Pure and Applied
# Mathematics

# Intersection Graphs: An Introduction

## *Madhumangal Pal*

Department of Applied Mathematics with Oceanology and Computer Programming
Vidyasagar University, Midnapore – 721102, India
e-mail: mmpalvu@gmail.com

**Abstract.** Intersection graphs are very important in both theoretical as well as application point of view. Depending on the geometrical representation, different type of intersection graphs are defined. Among them interval, circular-arc, permutation, trapezoid, chordal, disk, circle graphs are more important. In this article, a brief introduction of each of these intersection graphs is given. Some basic properties and algorithmic status of few problems on these graphs are cited. This article will help to the beginners to start work in this direction. Since the article contains a lot of information in a compact form it is also useful for the expert researchers too.

**Keywords:** Design and analysis of algorithms, intersection graphs, interval graphs, circular-arc graphs, permutation graphs, trapezoid graphs, tolerance graphs, chordal graphs, circle graphs, string graphs, disk graphs

***AMS Mathematics Subject Classification (2010):*** 05C62, 68Q22, 68Q25, 68R10

## 1. Introduction
It is well known that graph is a very useful tool to model problems originated in all most all areas of our life. The geometrical structure of any communication system including Internet is based on graph. The logical setup of a computer is designed with the help of graph. So graph theory is an old as well as young topic of research. Depending on the geometrical structures and properties different type of graphs have emerged, viz. path, cycle, complete graph, tree, planar graph, chordal graph, perfect graph, intersection graph, etc.

In this article, we concentrate our discussion on intersection graphs.

Suppose $S = \{S_1, S_2, \ldots\}$ be a set of sets. Draw a vertex ($v_i$) for each set $S_i$ and two vertices $v_i$ and $v_j$ are joined by an edge if the corresponding sets have a non-empty intersection, i.e. the edge $E$ is given by $E = \{(v_i, v_j) : S_i \cap S_j \neq \phi\}$.

An undirected graph $G = (V, E)$ is said to be $\chi -$ perfect if $\omega(G(A))$ $= \chi(G(A))$, for all $A \subseteq V$, and $G$ is said to be $\alpha -$ perfect if $\alpha(G(A)) = \kappa(G(A))$, for all $A \subseteq V$ where $G(A)$ is a subgraph induced by a subset $A$ of vertices. A graph is called *perfect* if it is either $\chi -$ perfect or $\alpha -$ perfect. It was proved in the famous Perfect Graph Theorem that a graph is $\chi -$ perfect if and only if it is $\alpha -$ perfect.

An undirected graph $G$ is called *p-critical* if it is minimally imperfect, i.e. $G$ is not perfect but every proper induced subgraph of $G$ is a perfect graph.

An undirected graph $G$ is called *triangulated* if every cycle of length strictly greater than three possesses a chord. In the literature, triangulated graphs are also called as *chordal, rigid-circuit, monotone transitive* and *perfect elimination* graphs.

The *clique graph* $C(G)$ of a graph $G$ is the intersection graph of the family of all cliques of $G$. The intersection graphs of conformal hypergraphs are just the clique graphs. *Cographs* (also called complement reducible graphs) are defined as the graphs which can be reduced to single vertices by recursively complementing all connected subgraphs.

For $Y \subseteq V$ let $G(Y)$ be the subgraph induced by $Y$. A vertex $v$ is *simplicial* if and only if $N[v]$ is a clique. The ordering $(v_1, v_2, \ldots, v_n)$ of $V$ is a perfect elimination ordering if and only if for all $i \in \{1, 2, \ldots, n\}$ the vertex $v_i$ is simplicial in $G_i$, where $G_i = G(\{v_i, v_{i+1}, \ldots, v_n\})$. The graph $G$ is *chordal* if and only if $G$ has a perfect elimination ordering. The ordering $(v_1, v_2, \ldots, v_n)$ is a strong elimination ordering if and only if for all $i \in \{1, 2, \ldots, n\}$, $N_i[v_j] \subseteq N_i[v_k]$ when $v_j, v_k \in N_i[v_i]$ and $j < k$ where $N_i[v]$ $(N_i(v))$ is the closed (open) neighbourhood of $v$ in $G_i$. The graph $G$ is *strongly chordal* if and only if $G$ has a strong elimination ordering. A vertex $u \in N[v]$ is a maximum neighbourhood of $v$ if and only if for all $w \in N[v]$ the inclusion $N[w] \subseteq N[u]$ holds (here $u = v$ is not excluded). The ordering $(v_1, v_2, \ldots, v_n)$ is a maximum neighbourhood ordering if for all $i \in \{1, 2, \ldots, n\}$ there is a maximum neighbour $u_i \in N_i[v_i]$, for all $w \in N_i[v_i]$, $N_i[w] \subseteq N_i[u_i]$ holds. The graph $G$ is *dually chordal* if and only if $G$ has a maximum neighbourhood ordering. The graph $G$ is *doubly chordal* if and only if $G$ is chordal and dually chordal. There is a close connection between chordal and dually chordal graphs which can be expressed in terms of *hypergraphs*.

Let $G = (V, E)$ be a graph. A set $S \subseteq V$ is an *asteroidal set* if for every $x \in S$ the set $S - \{x\}$ is contained in one component of $G - N[x]$. An asteroidal set with three vertices is called an *asteroidal triple (AT)*. A graph $G$ is called AT-free if $G$ has no AT. It may be noted that every asteroidal set is an independent set. A triple $\{x, y, z\}$ of vertices of $G$ is an AT if and only if for every two of these vertices there is a path between them avoiding the closed neighbourhood of the third.

A graph is a *comparability graph* if its edges can be given a transitive orientation. A *cocomparability* graph is the complement of a comparability graph. In the next subsection, some intersection graphs are defined some of them belong to comparability graphs where as some belong to cocomparability graphs.

## 1.1. Intersection graphs

The theory of intersection graphs have, together with others, its own mathematics classification number 05C62. This promotion may be mainly due to the fact that intersection graphs have nice applications in reality. A lot of properties of intersection graphs are available in [60]. We define intersection graph as follows.

A graph $G = (V, E)$ is called an *intersection graph* for a finite family $F$ of a non-empty set if there is a one-to-one correspondence between $F$ and $V$ such that two sets in $F$ have non-empty intersection if and only if their corresponding vertices in $V$ are adjacent. We call $F$ an intersection model of $G$. For an intersection model $F$, we use $G(F)$ to denote the intersection graph for $F$.

Depending on the nature or geometric configuration of the sets $S_1, S_2, \ldots$ different types of intersection graphs are generated. The most useful intersection graphs are

- Interval graphs ($S$ is the set of intervals on a real line)
- Tolerance graphs
- Circular-arc graphs ($S$ is the set of arcs on a circle)
- Permutation graphs ($S$ is the set of line segments between two line segments)
- Trapezoid graphs ($S$ is the set of trapeziums between two line segments)
- Disk graphs ($S$ is the set of circles on a plane)
- Circle graphs ($S$ is the set of chords within a circle)
- Chordal graphs ($S$ is the set of connected subgraphs of a tree)
- String graphs ($S$ is the set of curves in a plane)
- Graphs with boxicity $k$ ($S$ is the set of boxes of dimension $k$)
- Line graphs ($S$ is the set of edges of a graph).

It is interesting that every graph is an intersecting graph. For each vertex $v_i$ of $G$, form a set $S_i$ consisting of edges incident to $v_i$, the two such sets have a non-empty intersection if and only if the corresponding vertices share an edge.

An example of intersection graph is depicted in Figure 1. Let $S_+ = \{1, 2, 3, \ldots\}$, $S_- = \{-1, -2, -3, \ldots\}$, $S_p = \{2, 3, 5, 7, 11, \ldots\}$, $S_f = \{1, 1, 2, 3, 5, 8, \ldots\}$, $S_e = \{0, \pm2, \pm4, \ldots\}$, $S_o = \{\pm1, \pm3, \ldots\}$ be the sets of positive integers, negative integers, primes, Fibonacci numbers, even integers and odd integers respectively.

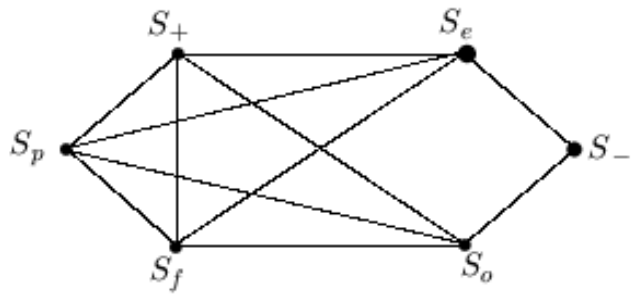

**Figure 1:** An intersection graph

Different types of intersection graphs are discussed in the subsequent sections.

## 2. Interval Graphs

An undirected graph $G = (V, E)$ is said to be an *interval graph* if the vertex set $V$ can be put into one-to-one correspondence with a set $I$ of intervals on the real line such that two vertices are adjacent in $G$ if and only if their corresponding intervals have non-empty intersection. That is, there is a bijective mapping $f : V \rightarrow I$.

The set $I$ is called an interval representation of $G$ and $G$ is referred to as the interval graph of $I$ [42].

Interval graphs arise in the process of modelling many real life situations, specially involving time dependencies or other restrictions that are linear in nature. This graph and various subclass thereof arise in diverse areas such as archeology, molecular biology, sociology, genetics, traffic planning, VLSI design, circuit routing, psychology, scheduling, transportation, etc. Also, interval graphs have found applications in protein sequencing [56], macro substitution [73], circuit routine [74], file organization [19], job scheduling [19], routing of two points nets [49] and so on. An extensive discussion of interval graphs is available in [42]. In addition to these, interval graphs have been studied intensely from both the theoretical and algorithmic point of view.

In the following an application of interval graph to register allocation is presented.

A computer program stores the values of its variables in memory. For arithmetic computations, the values must be entered in easily accessed locations called registers. Registers are expensive, so we want to use them efficiently. If two variables are never used simultaneously, then we can allocate them to the same register, one after use of other. For each variable, we compute the first and last time when it is used. A variable is active during the interval between these times.

We define a graph whose vertices are the variables. Two vertices are adjacent if they are active at a common time. The number of registers needed is the chromatic number of this graph. The time when a variable is active is an interval, so we obtain a special type of representation called interval graph. Let us consider a program segment shown in Figure 2(a).

The corresponding interval representation and interval graph are shown in Figure 2(b) and 2(c) respectively. Note that the chromatic number of the graph of Figure 2(c) is 4. That is, to execute this program segment of Figure 2(a), only 4 registers are required. Figure 2(d) shows the allocation of registers/colours. Note that an interval graph can be coloured in $O(n)$ time, where $n$ is the number of vertices [83].

Let us consider another example.

*Suppose a company or an organization is interested to run its advertisement through television channels. The constraint is that only one programme slot is selected for advertisement at any instant. The objective of the problem is to select the programme slots such that the sum of the number of viewers of the selected programmes is maximum.*

In this problem, a television programme slot is represented as a subinterval of an interval of length of 24 h on the real line. Each programme slot, i.e. each interval is regarded as a vertex of the interval graph $G$. That is, all the programme slots of all television channels constitute the set of vertices $V$ of the interval graph $G$. If there exists an intersection of timings in between two programme slots, there is an edge between the vertices corresponding to these programme slots. If the finishing time of a programme is the starting time of another programme then we assume that these programmes are non-intersecting. It may be noted that any two programmes in a particular television channel are always non-intersecting. The number of viewers of each programme is considered as the weight of the corresponding vertex of the graph. All the programme slots of all television channels are modelled as an interval graph.

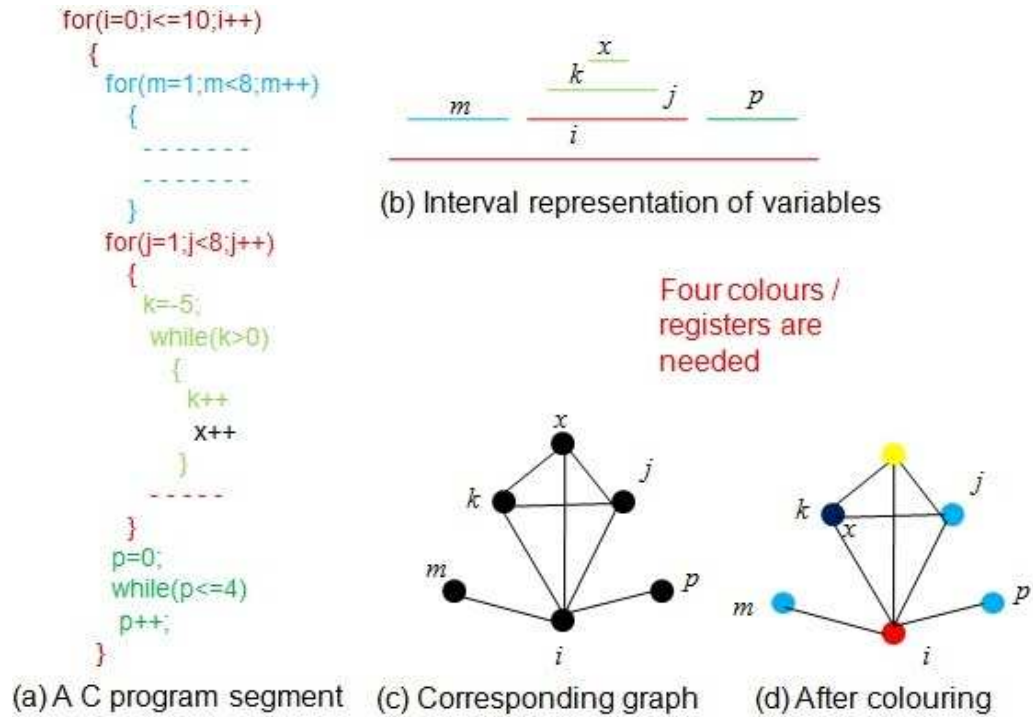

**Figure 2:** An application of interval graph in register allocation

A colouring of a graph is an assignment of colours to its vertices so that no two adjacent vertices have the same colour. The vertices of same colour form a colour class. Any two vertices of a colour class are not adjacent. The maximum weight colouring problem is to find a subset $S$ of $V$ such that no two vertices of $S$ are adjacent and sum of the weights of the vertices of $S$ is maximum. This problem is also known as maximum weight 1-colouring problem. Thus, the above problem is equivalent to the 1-colouring problem on interval graph. As mentioned above, all the programme slots of all channels in a geographical area can be represented as an interval graph. The details of this problem are described and solved in [105].

Interval graphs satisfy a lot of interesting properties. The first one is the *hereditary property*.

**Lemma 2.1.** *An induced subgraph of an interval graph is an interval graph [42].*

The next property of interval graphs is also a hereditary property, called *triangulated graph property*, which is stated below.

*Every simple cycle of length strictly greater than 3 possesses a chord.*

The graphs which satisfy this property are called triangulated graphs. So we have the following lemma.

**Lemma 2.2.** *An interval graph satisfies the triangulated graph property [46].*

Another important property on graphs is transitive orientation property stated below:

Each edge can be assigned a one-way direction in such a way that the resulting oriented graph $(V, E)$ satisfies the following condition:

$$(u, v) \in E \text{ and } (v, w) \in E \Rightarrow (u, w) \in E, \, u, v, w \in V.$$

The following result is due to Ghouila-Houri [38].

**Lemma 2.3.** *The complement of an interval graph satisfies the transitive orientation property.*

Let $G = (V, E)$ be a graph. A set of vertices $C \subseteq V$ forms a *clique* in $G$ if every pair of vertices in $C$ are adjacent. A *maximal clique* is a clique to which no further vertex of the graph can be added so that it remains a clique. A *maximum clique* is a clique with maximum cardinality.

The following theorem posed by Gilmore and Hoffman [41] establishes the position of the interval graphs in the world of perfect graphs.

**Theorem 2.1.** *Let $G$ be an undirected graph. The following statements are equivalent.*

*(i) $G$ is an interval graph.*

*(ii) $G$ contains no chordless cycle of length 4 and its complement $\overline{G}$ is a comparability graph.*

*(iii) The maximal cliques of $G$ can be linearly ordered such that, for every vertex $u$ of $G$, the maximal cliques containing $u$ occur consecutively.*

Statement (iii) of this theorem has an interesting matrix formulation. A matrix whose entries are zeros and ones, is said to have the *consecutive 1's property for columns* if its rows can be permuted in such way that the 1's in each column occur consecutively.

All maximal cliques of an interval graph can be computed in $O(n)$ time [76]. The maximal cliques versus vertices incidence matrix of a graph $G$ is called *clique matrix*.

The following theorem given by Fulkerson and Gross [34], is useful to recognize an interval graph.

**Theorem 2.2.** *An undirected graph $G$ is an interval graph if and only if its clique matrix $M$ has the consecutive 1's property for columns.*

Another important characterization of interval graph proposed by Lekkerkerker and Boland [58] is given below.

**Theorem 2.3.** *An undirected graph $G$ is an interval graph if and only if the following two conditions are satisfied:*

*(i) $G$ is a triangulated graph and*

*(ii) any three vertices of $G$ can be ordered in such a way that every path from the first vertex to the third vertex passes through a neighbour of the second vertex.*

The necessary and sufficient condition that a graph is an interval graph is stated below:

**Theorem 2.4.** [58] *A graph is an interval graph if and only if it contains none of the graphs shown in Figure 3 as an induced subgraph.*

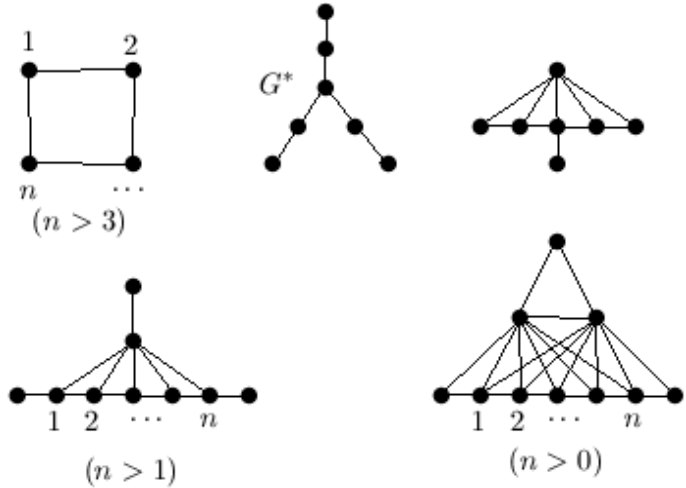

**Figure 3:** Forbidden structures for interval graphs

**Corollary 2.1.** *A tree is an interval graph if and only if it does not contain $G^*$ (Figure 3) as an induced subgraph.*

Let $G = (V, E)$, where $|V| = n, |E| = m$ be a simple connected interval graph, where vertices are numbered as $1, 2, \ldots, n$. Let $I = \{I_1, I_2, \ldots, I_n\}$ be the interval representation of an interval graph $G$, where $a_r$ is the left end point and the $b_r$ is the right endpoint of the interval $I_r$, i.e. $I_r = [a_r, b_r]$ for all $r = 1, 2, \ldots, n$. Without any loss of generality we assume the following:

- The intervals in $I$ are indexed by increasing right endpoints, i.e. $b_1 < b_2 < \cdots < b_n$.

- The intervals are closed, i.e. contains both of its endpoints and that no two intervals share a common endpoint.

- Vertices of the interval graph and the intervals on the real line are one and the same thing.

- The interval graph $G$ is connected and the list of sorted endpoints is given.

It is shown, in [91], that the set intervals of every interval graph can be ordered in a non-decreasing order of their right endpoints and this ordering is referred as IG ordering. In this section, the vertices are labelled as IG ordering. The IG ordering is obviously unique when a representation by a set of intervals is provided and fixed.

The following lemma is a powerful result on interval graph.

**Lemma 2.4.** *If the vertices $u, v, w \in V$ are such that $u < v < w$ in the IG ordering and $(u, w) \in E$, then $(v, w) \in E$.*

An interval graph and its interval representation are illustrated in Figure 4.

A lot of algorithms have been designed for interval graphs using different techniques. A very useful data structure called *interval tree* (IT) is defined in [84] which is discussed below.

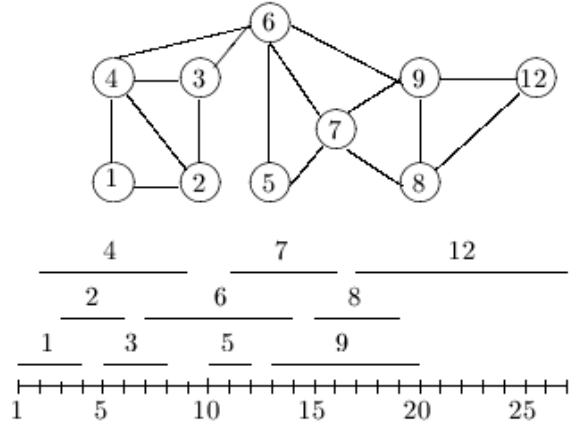

**Figure 4:** An interval graph and its interval representation.

**Interval tree**

For each vertex $v \in V$ let $H(v)$ and $L(v)$ represent respectively the highest and the lowest numbered adjacent vertices of $v$. It is assumed that $(v, v) \in V$ is always true. So, if no adjacent vertex of $v$ exist with higher (or lower) IG order than $v$ then $H(v)$ (or $L(v)$) is assumed to be $v$. In other words,

$$H(v) = \max\{u : (u, v) \in E, u \geq v\}, \ \ and$$
$$L(v) = \min\{u : (u, v) \in E, u \leq v\}.$$

It may be observed that the array $H$ is monotonic non-decreasing, i.e. if $u, v \in V$ and $u < v$ then $H(u) \leq H(v)$.

For a given interval graph $G$ let a spanning subgraph $G' = (V, E')$ be defined as
$$E' = \{(u, v) : u \in V \ and \ v = H(u), u \neq n\}.$$
The subgraph $G'$ of a connected interval graph $G$ is a tree [84]. Since the subgraph $G'$ is built from the vertex set $V$ and the edge set $E'$, where $E' \subseteq E$, $G'$ is a spanning tree of $G$. In what follows the subgraph $G'$ is referred to as *interval tree* and it is denoted by $T_I(G)$. In [84], it is shown that the interval tree $T_I(G)$ of a connected interval graph $G$ exists and is unique for a given interval representation.

The interval tree $T_I(G)$ of the interval graph of Figure 4 is shown in Figure 5. The level of a vertex $u$ in the interval tree is denoted by $level_I(u)$. Let $N_i$ be the set of vertices which are at a distance $i$ from the vertex $n$, i.e. $N_i$ is the set of vertices at level $i$. Thus $N_i = \{u : \delta_G(u, n) = i\}$ and $N_0$ is the singleton set $\{n\}$. It may be noted that if $u \in N_i$ then $level_I(u) = i$. Let $k$ be the maximum length of a shortest path from the vertex $n$ to any other vertex in $G$. It is easy to see that $N_k$ is non-empty while $N_{k+1}$ is empty.

## 2.1. Properties of interval tree

Let $\min(N_i)$ and $\max(N_i)$ represent the minimum and maximum vertices of the set

$N_i$ . That is, $\min(N_i) = \min\{u : u \in N_i\}$ and $\max(N_i) = \max\{u : u \in N_i\}$ . The vertices of $N_i$ are consecutive integers and $\max(N_{i+1}) = \min(N_i) - 1$ for all $i$ .

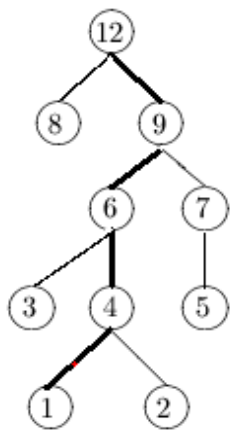

**Figure 5:** The interval tree of the graph of Figure 4.

It is easy to verify that if $level_I(u) < level_I(v)$ then $u > v$ .

The maximum value of $level_I(v)$ is $h(T_I(G))$ and the minimum value of $level_I(v)$ is 0. This minimum occurs when $v = n$ . But, if $v = n$ then $d(u,n) = level_I(u) \le h(T_I(G))$ . Thus, $\delta_G(u,v)$ is maximum when $level_I(v) = 1$ and $level_I(u) = h(T_I(G))$ and the maximum distance is $h(T_I(G)) + 1$ . From the definition of label it follows that $level_I(1) = h(T_I(G))$ and $level_I(n) = 0$ .

It is easy to note that the path from the vertex 1 to the vertex $n$ in the interval tree $T_I(G)$ is the longest path among the paths ending at $n$ . This path is referred as *main path*. The main path of the graph of Figure 4 is shown by thick lines (see Figure 5).

If two vertices have same level then the distance in $G$ between them is either 1 or 2. This result is given in the following lemma.

**Lemma 2.5.** [85] *For $u,v \in V$ if $level_I(u) = level_I(v)$ then*

$$\delta_G(u,v) = \begin{cases} 1, & \textit{if } (u,v) \in E(G) \\ 2, & \textit{otherwise.} \end{cases}$$

But, if $level_I(u) = level_I(v), u,v \in V$ then $\delta_{T_I(G)}(u,v)$ is not necessarily 1 or 2, it may even be more than 3 units. For example, for the interval graph of Figure 5, $level_I(3) = level_I(5) = 3$ and $\delta_{T_I(G)}(3,5) = 4$ .

If level of the vertex $v$ is $j$ then it should be adjacent in $G$ only to the vertices at level $j-1$ , $j$ and $j+1$ . If $u,v \in V$ and $|level_I(v) - level_I(u)| > 1$ then $(u,v) \notin E(G)$ [85].

Using the properties of interval tree lot of problems have been solved, some of them are listed below.

## 2.2. Applications of interval tree
## Construction of tree 3-spanner

A $t$-spanner of a graph $G$ is a spanning subgraph $H(G)$ in which the distance between every pair of vertices is at most $t$ times their distance in $G$, i.e. $\delta_H(u,v) \leq t \, \delta_G(u,v)$, for all $u,v \in V$. The parameter $t$ is called the *stretch factor*.

The *minimum $t$-spanner* problem is to find a $t$-spanner $H$ with the fewest possible edges for fixed $t$. The spanning subgraph $H$ is called a minimum $t$-spanner of $G$ and it is denoted by $H_t(G)$. A *spanning tree* of a connected graph $G$ is an acyclic (cycle free) connected spanning subgraph of $G$. A *tree spanner* of a graph is a spanning tree that approximates the distance between the vertices in the original graph. In particular, a spanning tree $T$ is said to be a *tree $t$-spanner* of a graph $G$ if the distance between any two vertices in $T$ is at most $t$ times their distance in $G$, i.e. $\delta_T(u,v) \leq t \, \delta_G(u,v)$ for all $u,v \in V$. It is obvious that if $G$ is connected, then $|E(H_t(G))| \geq n-1$, equality holds if and only if $G$ admits a tree $t$-spanner.

The optimization version of this problem is to find a $t$-spanner with the fewest possible edges for a fixed $t$.

It can be shown by examples that the interval tree may or may not be a tree 3-spanner of the corresponding interval graph.

**Lemma 2.6.** *The interval tree may or may not be a tree 3-spanner.*

The tree 3-spanner, $T_{3S}(G)$, of $G$ can be constructed from the interval tree by rearranging the parent vertex of some vertices. The method is described below:

Let $w_l^*$ and $w_{l+1}^*$ be two vertices on the main path at levels $l$ and $l+1$ respectively. Then assign parent of each vertex $u$, satisfying $w_{l+1}^* < u < w_l^*$ as $u_l^*$, i.e. $parent_I(u) = w_l^*$, where $parent_I(u)$ is the parent of the vertex $u$ in the interval tree $T_I(G)$. This process is repeated for all vertices of all levels $l$, $l = 1,2,\ldots,h(T_I(G))-1$. In other words, if $N_l = \{x_1,x_2,\ldots,x_{i-1},w_l^*,x_{i+1},\ldots,x_p\}$ and $N_{l+1} = \{y_1,y_2,\ldots,y_{j-1},w_{l+1}^*,y_{j+1},\ldots,y_q\}$, where $p = |N_l|$ and $q = |N_{l+1}|$ then parent of $y_{j+1},\ldots,y_q$ and $x_1,x_2,\ldots,x_{i-1}$ are $w_l^*$.

If $N_l = \{x_1,x_2,\ldots,x_{i-1},w_l^*,x_{i+1},\ldots,x_p\}$, where $p = |N_l|$, then let $N_l' = \{x_1,x_2,\ldots,x_{i-1}\}$ and $N_l'' = \{x_{i+1},\ldots,x_p\}$. That is, $N_l'$ (respectively, $N_l''$) is the subset of $N_l$ whose vertices are less than $w_l^*$ (respectively, greater than $w_l^*$). The set of vertices at level $l$ of the tree $T_{3S}(G)$ is denoted by $N_l^*$. Then from the construction of $T_{3S}(G)$, $N_{l+1}^* = \{w_{l+1}^*\} \cup N_{l+1}'' \cup N_l'$, $l = 0,1,\ldots,h(T_I(G))-1$.

The interval tree exits and is unique for a given interval representation of an interval graph. Also, the tree 3-spanner $T_{3S}(G)$ is obtained by rearranging the parents of the vertices of interval tree $T_I(G)$. Thus, every connected interval graph has a tree 3-spanner and it is unique for a given interval representation.

**Theorem 2.5.** [101] *A tree 3-spanner of an interval graph with $n$ vertices can be constructed in $O(n)$ time, if the sorted intervals are given.*

**Computation of diameter**

Let $G = (V, E)$ be a graph and $\delta_G(u, v)$ be the shortest distance between the vertices $u$ and $v$. The eccentricity of the vertex $u$ is denoted by $ecen(u)$ and is defined as

$$ecen(u) = \max_{v \in V} \{ \delta_G(u, v) \}.$$

The radius ($\rho(G)$) and diameter ($diam(G)$) of a graph $G$ are defined as

$$\rho(G) = \min_{u \in V} \{ ecen(u) \}, \ \ diam(G) = \max_{u \in V} \{ ecen(u) \}.$$

The diameter of an interval graph $G$ and the height of the corresponding interval tree $T_I(G)$ of $G$ are related by the following relation.

**Lemma 2.7.** [77] *Let $v_1^* \in N_1$ be the vertex on the main path. If all $v_1 \in N_1$ are adjacent to $v_1^*$ in $G$ then $diam(G) = h(T_I(G))$, otherwise $diam(G) = h(T_I(G)) + 1$.*

**Theorem 2.6.** *The diameter and centre of an interval can be computed in $O(n)$ time [78,84].*

**All-pairs shortest distances**

According to Lemma 2.5, the shortest distance between the vertices $u$ and $v$, when $level_I(u) = level_I(v)$, is either 1 or 2. But, if their levels are different then the distance between two vertices may be 1 or 2 or more. In this case, the distance between any two vertices can also be computed easily with the help of interval tree. The technique is described below.

To compute the distance between $u$ and $v$, $u < v$, we have to check the adjacency of the vertex $v$ with the vertices at levels $level(v) + 1$, $level(v)$ and $level(v) - 1$. Hence the distance $\delta_G(u, v)$ between any two vertices $u, v \in V$ can be computed using the following lemma.

**Lemma 2.8.** [85] *Given $u, v \in V$, let $z_1$ be the vertex at level $level(v) + 1$ on the path marked $\min(u)$ and $z_2 = H(z_1)$. If $level(u) > level(v)$, then*

$$\delta_G(u, v) = \begin{cases} level(u) - level(v), & if \ (z_1, v) \in E \\ level(u) - level(v) + 1, & if \ (z_1, v) \notin E \ and \ (z_2, v) \in E \\ level(u) - level(v) + 2, & otherwise \end{cases}$$

Using the above lemma the all-pairs shortest distances can be computed for an interval graph. The time complexity is presented below.

**Theorem 2.7.** [85] *The all-pairs shortest distances of an interval graph with $n$ vertices can be computed in $O(n^2)$ time.*

### $k$-neighbourhood covering problem

A vertex $x, k$-dominates another vertex $y$ if $d(x,y) \leq k$. A vertex $z, k$-neighbourhood-covers (kNC) an edge $(x,y)$ if $d(x,z) \leq k$ and $d(y,z) \leq k$, i.e. the vertex $z$ $k$-dominates both $x$ and $y$. Conversely, if $d(x,z) \leq k$ and $d(y,z) \leq k$ then the edge $(x,y)$ is said to be kNC by the vertex $z$. A set of vertices $C \subseteq V$ is a kNC set if every edge in $E$ is kNC by some vertex in $C$. The kNC number $\rho(G,k)$ of $G$ is the minimum cardinality of all kNC sets of $G$.

The kNC problem is a variant of the domination problem. Hwang and Chang [55] proved that kNC problem is NP-complete for general graph even for chordal graph. Mondal et al. [68] have solved 2-neighbourhood-covering problem in $O(n)$ time on interval graphs. Recently, Ghosh and Pal [40] have solved 2-neighbourhood-covering problem in $O(n)$ time on trapezoid graphs. The edge-paking problem is a kind of domination problem and this problem can be solved in $O(n)$ time on interval graphs [80].

By constructing a suitable interval tree for a given interval graph the kNC problem has been solved in [8].

**Theorem 2.8.** [8] *The kNC set of an interval graph can be computed in $O(n)$ time using $O(n)$ space, where $n$ is the number of vertices of the graph.*

### Conditional covering problem

The conditional covering problem (CCP in short) is a facility location problem on a graph [94]. Let $G = (V, E)$ be a graph where $V = \{1, 2, \ldots, n\}$ is the set of vertices and $E$ is the set of edges. The vertex set $V$ of the graph represents the set of demand points as well as the set of potential facility locations. A weight $w(e)$ is associated with every edge $e \in E$. The length of a path is the sum of the weights of the edges in the path. A path from the vertex $x$ to the vertex $y$ is a shortest path if there is no other path from $x$ to $y$ with lower length. We use $\delta(x,y)$ to denote such a shortest path. For a facility located at a vertex $x \in V$ requires a facility location cost $c(x)$ and provide a positive coverage radius $R(x)$. A facility can cover all vertices within its coverage radius except the vertex at which it is located, that is, a vertex $x \in V$ covered by a facility located at a vertex $y \in V$ if $x \neq y$ and $\delta(x,y) \leq R(x)$. The CCP seeks to minimize facility location cost such that the set of vertices in the graph must be covered by a facility and every facility should be covered by at least one another facility. Here each facility has a specified, possibly overlapping region to serve.

One closely related to the CCP is the total dominating set problem, which is a special case of the CCP in which all distances and coverage radii equal 1. It should be noted that if $R = 1$, then our problem is identical to the *total dominating set* problem. The total dominating set problem is NP-complete even for bipartite graph [50]. Since the total dominating set is a special case of CCP, the CCP is also NP-complete for general graphs.

**Theorem 2.9.** [92] *The time complexity to find the conditional covering set on unweighted cost interval graph is* $O(n)$.

**Theorem 2.10.** [93] *The CCP on interval graph with non-uniform cost can be solved in* $O(n^2)$ *time.*

**Minimum feedback vertex problem**

Given $S \subseteq V$, the induced subgraph $G(S)$ is the graph $G(S) = (S, E(S))$, where $E(S) = \{(v_i, v_j) : v_i, v_j \in S, (v_i, v_j) \in E\}$. A set $S$ is a feedback vertex set if and only if the graph $G(V - S)$ has no cycles. The minimum feedback vertex set (MVFS) problem is to find a feedback vertex set $S$ such that its cardinality $|S|$ is minimum among all such sets.

The minimum feedback vertex set problem is NP-hard for general graphs.

**Theorem 2.11.** [102] *The minimum cardinality feedback vertex set of an interval graph with* $n$ *vertices and* $m$ *edges can be computed in* $O(n + m)$ *time.*

**Theorem 2.12.** [101] *The maximum weight feedback vertex set of a weighted interval graph can be computed in* $O(n\sqrt{\log C})$ *time, where* $C$ *is the weight of a longest path of the graph.*

Let $G = (V, E)$ be a simple undirected graph and $G(u)$ be the subgraph of it induced by the vertex set $V - \{u\}$. The distance $\delta_G(x, y)$ is defined to be the length of the shortest path between the vertices $x$ and $y$ in $G$. A vertex $u \in V$ is said to be a *hinge vertex* if there exist two vertices $x, y \in V - \{u\}$ such that $\delta_{G(u)}(x, y) > \delta_G(x, y)$. A graph without hinge vertices is called a *self repairing graph*.

**Theorem 2.13.** [12] *The set of all hinge vertices of an interval graph with* $n$ *vertices can be computed sequentially in* $O(n)$ *time using* $O(n)$ *space. Also, all hinge vertices of a trapezoid graph can be computed in O(n log n) time [14].*

The *vertex connectivity* (or simply *connectivity*) of a connected graph $G$ is defined as the minimum number of vertices whose removal from $G$ leaves the remaining graph disconnected or trivial. The vertex connectivity of a tree is thus one. A connected graph is said to be *separable* if its vertex connectivity is one. All other connected graphs are called non-separable. In a graph a vertex (or edge) whose removal disconnects the graph is called a cut vertex (or bridge) or a cut-node or an articulation point. It can be easily shown that in a tree every vertex with degree greater than one is a cut-vertex. In general: a vertex $v$ in a connected graph $G$ is a cut-vertex if and only if there exist two vertices $x$ and $y$ in $G$ such that every path between $x$ and $y$ passes through $v$. The maximal non-separable subgraphs of $G$ are called the blocks or biconnected components.

**Theorem 2.14.** [88] *All the cut-vertices and blocks of an interval graph can be computed in $O(n)$ time.*

### 2.3. $k$-gap interval graphs

A multiple interval representation $f$ of a graph $G = (V, E)$ is a mapping that assigns to each vertex of $G$ a non-empty collection of intervals on the real line, such that two distinct vertices $u$ and $v$ are adjacent if and only if there are intervals $I \in f(u)$ and $J \in f(v)$ with $I \cap J \neq \phi$. $|f(v)|$ denotes the number of intervals associated to $v$. The interval number of $G$ is defined as $i(G) = \min\{\max_{v \in V}\{|f(v)|\} : f$ is a multiple interval representation of $G\}$.

The total interval number of a graph $G = (V, E)$ is defined as $I(G) = \min\{\sum_{v \in V}\{|f(v)|\} : f$ is a multiple interval representation of $G\}$.

A new generalization of interval graphs called $t$-interval graphs is introduced by Trotter and Harary [110], and Griggs and West [45]. A graph $G$ is called $t$-interval graphs if $i(G) \leq t$. The $t$-interval graphs are applicable in scheduling and resource allocation [5], communication protocols [18], computational biology [24, 27], monitoring [18], etc.

The class of $t$-interval graph is richer than interval graphs, for example, the class of 2-interval graphs include circular-arc graphs, outerplanar graphs, cubic graphs, line graphs and 3-interval graphs include all planar graphs [106]. The class of graphs with maximum degree $\Delta$ are $\lfloor (\Delta+1)/2 \rfloor$-interval graphs [45], while the complete bipartite graph $K_{m,n}$ is a $\lfloor (mn+1)/(m+n) \rfloor$-interval graph. Every graph with $n$ vertices is a $\lfloor (n+1)/4 \rfloor$-interval graph.

**Definition 2.1.** *The $k$-gap interval graph is a graph that have a multiple interval representation whose total number of intervals exceed the number of vertices by at most $k$, i.e. a graph $G$ on $n$ vertices is a $k$-gap interval graph if $I(G) \leq n + k$.*

In a $k$-gap interval graph with multiple interval representation $f$, a vertex $v \in V$ has a gap if $|f(v)| \geq 2$.

A $k$-gap interval graph can be obtained from an interval graph by a sequence of at most $k$ operations of identifying pairs of vertices. A multiple interval representation of $G$ has $k$ gaps if $\sum_{v \in V} |f(v)| = |V| + k$.

**Definition 2.2.** *A graph $G = (V, E)$ is an interval $+ kv$ graph if there is a vertex set $X \subseteq V$, with $|X| \leq k$, such that $G \setminus X$ is an interval graph. The vertex set $X$ is called as interval deletion set of $G$.*

The recognition of $k$-gap interval graphs is NP-hard [57].

**Lemma 2.9.** [33] *An interval* $+ kv$ *graph on* $n$ *vertices has* $O(n2^k)$ *maximal cliques.*

Most of the problems, viz. recognition, 3-coloring, dominating set, independent sets, Hamiltonian cycle, etc. are NP-hard.

**Lemma 2.10.** [33] *Feedback vertex set and clique cover problems can be solved in* $(2^{O(k \log k)} n^{O(1)})$ *and* $O(2^k n^{O(1)})$ *times respectively, on interval* $+ kv$ *graphs with* $n$ *vertices.*

Several approximation algorithms are design to solve optimization problems on multiple-interval graphs [18].

## 2.4. Dotted interval graphs

A *dotted interval* $I(s,t,d)$ is an arithmetic progression $\{s, s+d, s+2d, \ldots, t_g\}$, where $s, t$ and $d$ are positive integers and $d$ is called the jump. When $d = 1$, the dotted interval $I(s,t,d)$ is simply the ordinary interval $[s,t]$ over the positive integer line. A *dotted interval graph* is an intersection graph of dotted intervals. Like interval graph, for each dotted interval $I_v$ there is a vertex $v$ and if $I_u \cap I_v \neq \phi$ then there is an edge between the vertices $u$ and $v$. If the jumps of all intervals are at most $d$, then the graph is called the $d$-dotted-interval ($d$-DIG). An example of 2-dotted interval graph is shown in Figure 6. In this figure we consider five dotted intervals $I_a = I(1,5,2) = \{1,3,5\}$, $I_b = I(2,3,1) = \{2,3\}$, $I_c = I(1,7,2) = \{1,3,5,7\}$, $I_d = I(4,6,2) = \{4,6\}$, $I_e = I(6,8,2) = \{6,8\}$. Note that each dotted interval is a set of integers and each set of integers forms an arithmetic progression.

In [3,4], Aumann et al. introduced the dotted interval graphs in the context of high throughput genotyping. They used dotted intervals to model microsatellite polymorphisms which are used in a genotyping technique called microsatellite genotyping. The respective genotyping problem translates to minimum coloring in $d$-DIGs for small $d$. They have shown that minimum coloring in $d$-DIGs is NP-hard even for $d = 2$.

**Theorem 2.15.** [3] *Every graph with a countable number of nodes is a dotted interval graph.*

**Theorem 2.16.** [3] *For all* $d \geq 1$, $d$ *-DIG* $\nsubseteq (d+1)$ *-DIG.*

**Theorem 2.17.** [47] *Maximum independent set and minimum vertex cover in* $d$ *-DIG graphs can be computed in* $O(dn^d)$ *-time.*

Using the approach to solve maximum independent set problem, the minimum dominating set problem in $d$-DIG graphs has been solved in [47].

**Theorem 2.18.** [47] *The minimum dominating set on* $d$ *-DIG graphs with* $n$ *vertices can be solved in* $O(d^2 n^{O(d^2)})$ *time.*

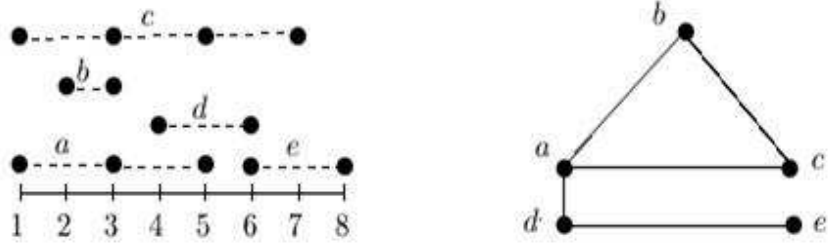

**Figure 6:** A set of dotted intervals and corresponding 2-dotted interval graph

It is well known that an interval graph with maximum clique size $k$ has pathwidth $k-1$.

**Definition 2.3.** *A clique $K$ in a $d$-DIG graph with dotted interval representation $I$ is a point clique if there exists a point $p \in N$ which is included in every $I_v \in I$ with $v \in K$.*

**Lemma 2.11.** [47] *A $d$-DIG graph with maximum point clique size $k$ has pathwidth at most $dk-1$.*

An interesting corollary of the above lemma is stated below.

**Corollary 2.2.** [47] *There is a linear-time algorithm for solving minimum vertex cover restricted to $d$-DIG graphs given with representations that have maximum point clique size $k$.*

**Theorem 2.19.** [3] *Coloring of $d$-DIGs is NP-complete for any $d \geq 2$.*

It may be noted that 1-DIGs are interval graphs, which are polynomially colorable. Thus, there is a clear distinction between 1-DIG and 2-DIG.

## 3. Tolerance graphs

Let $I = \{I_u : u \in V\}$ and $t = \{t_u : u \in V\}$ be the sets of intervals and tolerances. A graph $G = (V, E)$ is called a tolerance graph [43] if each vertex $v \in V$ can be assigned a closed interval $I_v$ and a tolerance $t_v \in R^+$ so that $(u, v) \in E$ if and only if $|I_u \cap I_v| \geq \min\{t_u, t_v\}$. Such a collection $\langle I, t \rangle$ of intervals and tolerances is called a tolerance representation. If graph $G$ has a tolerance representation with $t_v \leq |I_v|$ for all $v \in V$, then $G$ is called a bounded tolerance graph and the representation is called a bounded tolerance representation. Let

$I = \{I_a = [0,3], I_b = [1,4], I_c = [3,9], I_d = [8,11], I_e = [9,11], I_f = [4,8], I_g = [5,7]\}$

be a set of intervals and $t = \{t_a = 1, t_b = 1, t_c = 4, t_d = 1, t_e = 1, t_f = 1, t_g = \infty\}$ be a set of tolerances. The interval representation and the corresponding interval graph are shown in Figure 7.

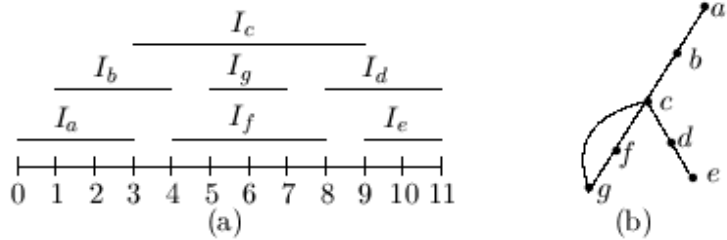

**Figure 7:** Interval representation and its corresponding interval graph

The tolerance representation $\langle I, t \rangle$ and the corresponding tolerance graph are shown in Figure 8.

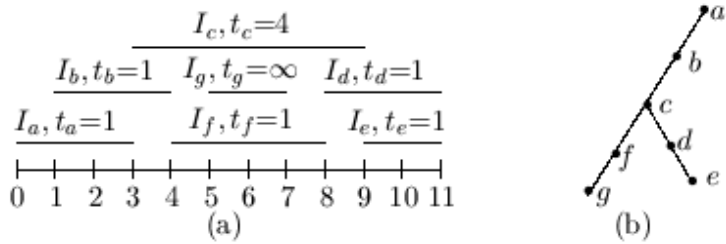

**Figure 8:** Tolerance representation and its corresponding tolerance graph

Note that there is no edge between the vertices $c$ and $g$ in tolerance graph (though $I_c \cap I_g \neq \phi$) since $|I_c \cap I_g| = 2 \not\geq \{t_c, t_g\}$.

That is, the graph of Figure 8(b) is a tolerance graph but not an interval graph.

It can be easily verified that an induced subgraph of a tolerance graph is a tolerance graph and induced subgraph of a tolerance graph and induced subgraph of a bounded tolerance graph is also a bounded tolerance graph. If $\langle I, t \rangle$ is a tolerance representation of a tolerance graph $G$ and $I_v$ is a point, then the corresponding vertex $v$ is an isolated vertex of $G$. Again, if $\langle I, t \rangle$ is bounded tolerance representation of $G$, then no interval $I_v$ is a point.

A tolerance representation is called regular [44] if

(i) any tolerance large than the length of its corresponding interval is set to infinity.

(ii) all tolerances are distinct.

(iii) no two different intervals share an endpoint.

**Lemma 3.1.** [44] *Every tolerance graph has a tolerance representation.*

If $t_v$ for all $v \in V$ are constant then the tolerances are called constant. For constant tolerance the following theorem is important.

**Theorem 3.1.** [43,44] *The following are equivalent statements about a graph $G$.*

*(i) $G$ is an interval graph.*

*(ii) $G$ is a tolerance graph with constant tolerance.*

*(iii) $G$ is a bounded tolerance graph with constant tolerances.*

Let $L_1$ and $L_2$ be two horizontal lines where $L_1$ is top line and $L_2$ is the bottom line. A graph $G$ is a *parallelogram graph* if each vertex $i \in V$ is associated to a parallelogram $P_i$ with parallel side along $L_1$ and $L_2$. Thus, $G$ is an intersection graph. This is also a special type of trapezoid graph.

**Theorem 3.2.** [44] *A graph is a bounded tolerance graph if and only if it is a parallelogram graph.*

In 1984, Golumbic, Monma and Trotter proved the following result.

**Theorem 3.3.** [44] *Tolerance graphs are weakly chordal.*

They also proved the following theorem.

**Theorem 3.4.** [44] *The tolerance graphs are perfect.*

A tolerance graph is called unit tolerance graph if all intervals have the same length. A graph is called 50% tolerance graph if it has a tolerance representation $\langle I, t \rangle$ so that $t_v = \frac{1}{2} |I_v|$ for all $v \in V$. In general, if $G$ has a tolerance representation for which there is a constant $c$ with $|I_v| - 2t_v = c$ for all $v \in V$, then the tolerance representation has constant cores.

**Theorem 3.5.** [44] *The following statements are equivalent.*
    *(i) $G$ is a unit tolerance graph.*
    *(ii) $G$ is a 50% tolerance graph.*
    *(iii) $G$ has a bounded tolerance representation with constant cores.*
It is easy to verify the following result.

**Theorem 3.6.** [16] *Interval graphs are unit tolerance graph.*

**Theorem 3.7.** [44] *Let $G = (X, Y, E)$ be a bipartite graph. The following conditions are equivalent*
    *(i) $G$ is a bounded tolerance graph.*
    *(ii) $G$ is a trapezoid graph.*
    *(iii) $G$ is a cocomparability graph.*
    *(iv) $G$ is AT-free.*
    *(v) $G$ is a permutation graph.*

## 4. Circular-arc graphs

A graph is a *circular-arc* graph if there exists a family $A$ of arcs around a circle and a one-to-one correspondence between vertices of $G$ and arcs in $A$, such that two distinct vertices are adjacent in $G$ if and only if the corresponding arcs intersect in $A$. Such a family of arcs is called an *arc representation* for $G$.

A graph $G$ is a *proper circular-arc (PCA)* graph if there exists an arc representation for $G$ such that no arc is properly included in another. Tucker [112] presented a characterization and an efficient recognition algorithm, using matrix characterizations, for recognizing PCA graphs. Deng *et al.* [29] presented a recognition algorithm that runs in linear time, and also produces a PCA model within this time bound when the graph is a PCA graph.

A graph $G$ is a *unit circular-arc (UCA)* graph if there exists an arc representation for $G$ such that all the arcs are of the same length. Tucker [113] presented a characterization by forbidden subgraphs for this class of graphs. This characterization shows that UCA graphs are proper subclass of PCA graphs. They can be useful in traffic control, when it is necessary that the green traffic lights for each lane at a street intersection are on for the same amount of time [42].

A family $S$ of subsets satisfied the Helly property when every subfamily of $S$ consisting of pairwise intersecting subsets has a common element. A graph is a *Helly circular-arc (HCA)* graph if there exists an arc model for $G$ such that the arcs satisfy the Helly property.

Circular-arc graphs have many applications in different fields such as genetic research, traffic control, computer compiler design, statistics, etc.. McConnell [65] proposed an $O(n+m)$ time algorithm for recognizing a circular-arc graph.

Let $S = \{a_1, a_2, \ldots, a_n\}$ be a family of $n$ arcs on a circle $C$. Each endpoint of the arcs is assigned a positive integer, called a *coordinate*. The endpoints of each arc are located on the circumference of $C$ in the ascending order of the values of the coordinates in the clockwise direction. For convenience, each arc $a_i, i = 1, 2, \ldots, n$, is represented as $(h_i, t_i)$, where $h_i$ (the *head*) and $t_i$ (the *tail*) denote, respectively the starting and ending points of the arc when it is traversed in clockwise manner, starting with an arbitrary chosen point on $C$ which is not an endpoint of any arc in $S$.

Without loss of generality, we assume the following:

1. No single arc in $S$ covers the entire circle $C$ by itself (otherwise, the shortest path problem becomes trivial and in this case the distance between any two arcs is either 1 or 2 unit).

2. No two arcs share a common endpoint.

3. $\bigcup_{i=1}^{n} C_i = C$ (otherwise, the problem becomes one on interval graph).

4. The endpoints of the arcs in $S$ are already given and sorted, according to the order in which they are visited during the clockwise (anticlockwise, if necessary) traversal along $C$ by starting at $a_1$.

5. The arcs are sorted in increasing values of $h_i$'s, i.e. $h_i > h_j$ for $i > j$.

The family of arcs $S$ is said to be *canonical* if

1. $h_i$'s and $t_i$'s for $i = 1, 2, \ldots, n$ are all distinct integers between 1 and $2n$, and

2. point 1 is the head of the arc $a_1$.

When $S$ is not canonical, using sorting one can construct a canonical family of arcs using $O(n \log n)$ time. In this article we assume that the family $S$ is canonical.

Given the set of arcs $S$, a *path* from the arc $a \in S$ to the arc $b \in S$ is a sequence $\sigma = (a_1, a_2, \ldots, a_k)$ of arcs in $S$ such that $a_1 = a$ and $a_k = b$, and $a_i$ and $a_{i+1}$ have some common points for every $i = 1, 2, \ldots, k-1$. The *length* of $\sigma$ is the number of arcs on $\sigma$, and $\sigma$ is a *shortest path* from $a$ to $b$ if it has the smallest length among all possible $a$-to-$b$ paths in $S$.

A circular-arc representation of an undirected graph $G$ which fails to cover some point $p$ on the circle will be topologically the same as an interval representation of $G$. In this case, we can cut the circle at $p$ and straighten it out to a straight line, the arcs become intervals. It is easy to see, therefore, that every interval graph is a circular-arc graph. The converse, however, is false.

For an arc $a_i \in S$ and an endpoint $j$ of another arc in $S$, we say that $a_i$ *contains point* $j$ if one of the following three conditions hold.

1. $1 \le h_i < j < t_i \le 2n$  (Figure 9(a)).
2. $1 \le t_i < h_i < j \le 2n$  (Figure 9(b)).
3. $1 \le j < t_i < h_i \le 2n$  (Figure 9(c)).

Let $G = (V, E)$ be a PCA graph and $\rho = \{a_1, a_2, \ldots, a_n\}$ a proper circular-arc model for $G$.

An $(n, k)$-*circuit* of $G$ with respect to $\rho$ is a set $\{v_1, v_2, \ldots, v_n\}$ of vertices ($n \ge 1$), such that $v_i$ and $v_{i+1}$ are adjacent ($1 \le i \le n-1$), $v_n$ is adjacent to $v_1$, arc $a_{i+1}$ starts clockwise from the counterclockwise endpoint of arc $a_i$, and the set of arcs wraps $k$ times around the circle. To count the number of turns around the circle, we walk along the circumference starting at the counterclockwise endpoint of $a_1$, jumping from $a_1$ to $a_2$, from $a_2$ to $a_3$, and so on, and each time we pass by the starting point, we count a new turn.

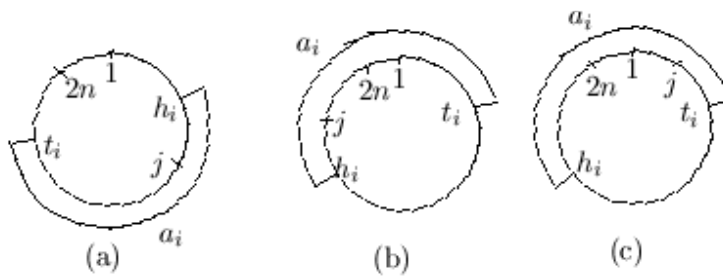

**Figure 9:** Position of the point $j$ within the are $a_i$ (a) $1 \le h_i < j < t_i \le 2n$, (b) $1 \le t_i < h_i < j \le 2n$, (c) $1 \le j < t_i < h_i \le 2n$

An $(m, l)$-*independent set* is a set $\{v_1, v_2, \ldots, v_m\}$ of vertices ($m \ge 1$), such that $v_i$ and $v_{i+1}$ are nonadjacent ($1 \le i \le m-1$), $v_m$ is nonadjacent to $v_1$, arc $a_{i+1}$ is an arc that starts clockwise from the clockwise endpoint of arc $a_i$, and the set of arcs wraps $l$ times around the circle. We count the number of turns around the circle in the same manner as before, but now we consider the last turn as complete, i.e. we add 1 to the total number of turns.

The graph $CI(n,k)$, with $n > k$, is the circular-arc graph corresponding to the circular-arc model built in the following way: let $\varepsilon$ be a positive real number and $r = 1$ be the radius of the circle. Draw $n$ arcs $a_0, a_1, \ldots, a_{n-1}$ of length $l_1 = 2\pi k/n + \varepsilon$ such that each $a_i$ starts at $2\pi i/n$ and ends at $2\pi(i+k)/n + \varepsilon$, i.e. $a_i = (2\pi i/n, 2\pi(i+k)$ $/n + \varepsilon)$. Afterward, draw $n$ arcs $b_0, b_1, \ldots, b_{n-1}$ of length $l_2 = 2\pi k/n - \varepsilon$, such that each $b_i$ starts at $(2\pi i + \pi)/n$ and ends at $(2\pi(i+k) + \pi)/n - \varepsilon$, i.e. $b_i = ((2\pi i + \pi)/n, (2\pi(i+k) + \pi)/n - \varepsilon)$.

**Theorem 4.1.** [113] *Let $G$ be a proper circular-arc graph. Then $G$ is a unit circular-arc graph if and only if $G$ contains no $CI(n,k)$ subgraphs, with $n, k$ relatively prime and $n > 2k$.*

**Theorem 4.2.** [31] *Let $G$ be a proper circular-arc graph. Then the following statements are equivalent:*

    *(i) $G$ is not a unit circular-arc graph,*

    *(ii) $G$ contains a $CI(n,k)$ subgraph, with $n, k$ relatively prime and $n > 2k$.*

The following theorem guarantees that every PCA graph can be represented with no circle-covering pairs of arcs.

**Theorem 4.3.** [42, 113] *If $G$ is a proper circular-arc graph, then $G$ has a proper circular-arc model in which no pair of arcs covers the circle.*

**Lemma 4.1.** [113] *If $G$ is a unit circular-arc graph, then it is also a proper circular-arc graph.*

For two distinct arcs $a_i$ and $a_j$ in $S$, we say that they *intersect* with each other if one of them contains at least one of the endpoints of the other arc; otherwise $a_i$ and $a_j$ are said to be *independent* from each other. If $a_i$ contains both endpoints of $a_j$, we say that $a_i$ *contains* $a_j$.

An undirected graph $G = (V, E)$ is a *circular-arc* graph if and only if

(i) its vertices are circularly indexed as $v_1, v_2, \ldots, v_n$, and

(ii) $(v_i, v_j) \in E$, provided $a_i$ and $a_j$ intersect with each other, where $v_i$ and $v_j$ are the vertices in the graph $G$ corresponding to the arcs $a_i$ and $a_j$ in $S$ respectively.

It may be noted that the arc $a_i$ and vertex $v_i$ or $i$ are one and the same thing.

**Corollary 4.1.** *A tree is a circular-arc graph if and only if it is an interval graph.*
**Proof:** Let $G$ be a circular-arc graph which is a tree and suppose that it is not an interval graph. Therefore, by Theorem 2.4, $G$ contain some of the graphs shown in Figure 3. Since $G$ is a tree, this induced subgraph can only be $G^*$. But this graph is not a circular-arc graph, which is a contradiction. The converse is true because interval graphs are a subclass of circular-arc graphs.                                                                ☐

The maximum independent set problem is NP-complete for general graphs [35]. We consider a fixed line drawn from the centre of the circle and passing through the finishing point of any arc. Some arcs are intersected by this line. This set of arcs is called the set of *backward arcs* denoted by $S_B$ and the remaining set of arcs are called the set of *forward arcs* denoted by $S_F$. Let $k$ be the number of arcs that are intersected by the line, i.e. $|S_B| = k$. First we label the endpoints of this set of arcs along the clockwise direction starting from 1 to $k$ as their successive starting points are encountered. Then we label the remaining arcs from $k+1$ to $n$ as their starting points are encountered. In this process, we specify an order of the arcs by increasing counterclockwise ends. If any two arcs $A_i$ and $A_j$ share the common points of the circle the arcs are said to be *intersecting arcs*. If two arcs do not share a common point of the circle then they are called *independent arcs*.

We are only interested in the case where every point on the circle covered by at least one arc. If some portions on the circle is not covered by any arc, of any circular-arc graph, then a gap exists. In this case, arc model of this graph can be viewed simply as an interval model that has been bent around the circle. In this case, the maximum weight independent set can be computed by applying any algorithm designed for interval graph.

In the following we present some results of circular-arc graph.

**Lemma 4.2.** *The arcs of $S_B$ form a clique.*

**Proof:** The arcs of $S_B$ must intersect each other at the line which is drawn from the circumference of circle, i.e. any two arcs intersect each other. Thus, they form a clique. ☐

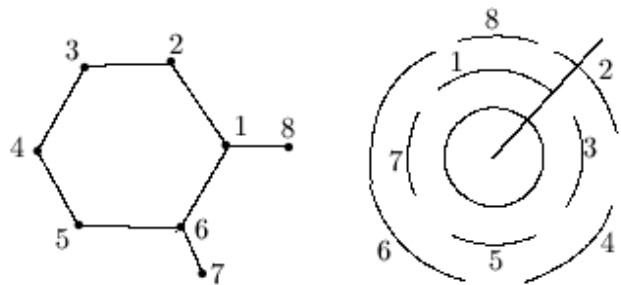

**Figure 10:** Example of a circular-arc graph and its circular arc representation.

**Lemma 4.3.** *If $X$ is the maximum weight independent set of $A$ and $Z$ be any clique, then $|X \cap Z| = 1$.*

**Proof:** Any two vertices of $Z$ are connected by at least one edge. So we cannot take more than one vertex of $Z$ into $X$ otherwise will not remain independent. Thus, $|X \cap Z| \leq 1$. Also, if $|X \cap Z| = \phi$ then including of any vertex of $Z$ into $X$, it remains independent and weight will be greater than the previous set. Hence, it must be $| X \cap Z |= 1$. □

The following lemma is very interesting.

**Lemma 4.4.** *The graph* $G \setminus N[i]$ *is an interval graph for any* $i$, *where* $N[i]$ *is the close neighbour of the vertex* $i$.

Let $S_i$ be the maximum weight independent set of the graph $G \setminus N[i]$. The set $S_i \bigcup \{i\}$ is a maximal independent set of the circular-arc graph $G$.

**Theorem 4.4.** [61] *The maximum weight independent set of a circular-arc graph with* $n$ *vertices can be computed using* $O(n^2)$ *time.*

**Theorem 4.5.** [63] *The set of hinge vertices of a circular-arc graph can be computed using* $O(n)$ *time, where* $n$ *is the number of vertices.*

The 2-neighbourhood covering problem on circular-arc graph is solved by converting it to an appropriate interval graph. The main reason for this conversion is that the interval graph can easily be take up with its good data structure interval tree. Pal and Bhattachajee [85] have developed the data structure interval tree. Thus, to solve the problem we first transfer the family of arcs to an equivalent family of intervals on a real line.

Let $A$ be the set of arcs of the circular-arc graph and $I$ be the set of intervals on the real line. First, we consider a ray through the finishing point of the arc $A_n$, i.e. $f_n$. We consider the arc $A_n$ as an interval $I_n$, where finishing endpoint of $A_n$ is right endpoint of $I_n$ and starting endpoint of $A_n$ is left endpoint of $I_n$. Similarly, we transfer all arcs $A_i, i = 1, 2, \ldots, n-1$ of the circular-arc graph $G$ to the interval $I_i$ of the interval graph. Also we add one more interval corresponding to the arc $A_n$ and we label this interval as 0. The left endpoint of the interval $I_0$ is less than the left endpoint of the interval $I_1$ and the right endpoint of $I_0$ is greater than the left endpoint of the interval $I_1$.

We define the interval graph corresponding to the circular-arc graph $G$ as $G' = (V', E')$. In $G'$, there is one more vertex corresponding to the interval $I_0$. So, we define the vertex set of the interval graph as $V'$ which is equal to $V \bigcup \{0\}$.

**Theorem 4.6.** *The 2-neighbourhood covering set of any one of the circular-arc graph can be computed in* $O(n)$ *time.*

To compute all-pairs shortest distances of the circular-arc graph the basic idea of our proposed algorithm is as follows. At first we transfer circular-arc graph into two interval graphs. Then we construct two interval trees for these two interval graphs. Next we find the distance between two vertices in each tree and minimum of those two values, i.e. two distances is the shortest distance between two vertices of the circular-arc graph.

**Theorem 4.7.** [103] *The all-pair shortest paths problem on circular-arc graph can be solved in* $O(n^2)$ *time.*

The next-to-shortest path from the vertex $i$ to the vertex $j$ is the shortest path from the vertex $i$ to the vertex $j$ amongst those the distances strictly greater than the shortest distance. If no such path exist, we say that the distance of next-to-shortest path is $\infty$. The length of the next-to-shortest path is called *next-to-shortest distance*.

**Theorem 4.8.** [62] *The next-to-shortest paths problem on circular-arc graph can be solved in $O(n^2)$ time.*

The *average distance* $\mu(G)$ of a connected graph is defined to be the average of all distances in $G$ as

$$\mu(G) = \frac{1}{n(n-1)} \sum_{\substack{x,y \in V \\ x \neq y}} \delta(x, y),$$

where $\delta(x, y)$ denotes the length of a shortest path joining the vertices $x$ and $y$. The average distance can be used as a tool in analytic networks where the performance time is proportional to the distance between any two nodes. It is a measure of the time needed in the average case, as opposed to the diameter, which indicates the maximum performance time.

**Theorem 4.9.** [64] *The average distance of a circular-arc graph can be computed in $O(n^2)$ time.*

## 5. Chordal graphs

A graph is said to be *chordal* if each of its cycles of four or more nodes has a chord, which is an edge joining two nodes that are not adjacent in the cycle. In other words, a chordal graph is a graph with no induced cycles of length more than three.

From the above definition one can say that an undirected graph is chordal if it does not contain an induced subgraph isomorphic to $C_n$ for $n > 3$. Thus, interval graphs are chordal.

Chordal graphs are a subset of perfect graphs [42]. They are also known as *rigid circuit graphs* or *triangulated graphs.*

The vertices of a chordal graph can be ordered in a systematic way which is very useful in designing algorithms on chordal graphs, is known as *perfect elimination ordering*. A vertex is called *simplicial* if its adjacency set induces a complete subgraph, that is, a clique (not necessarily maximal). A permutation $\sigma = [v_1, v_2, \ldots, v_n]$ of the vertices of an undirected graph $G$, or a bijection $\sigma : V \to \{1, 2, \ldots, n\}$, is called a perfect elimination order if each $v_i$ is a simplicial vertex of the subgraph of $G$ induced by $\{v_1, v_2, \ldots, v_n\}$.

**Theorem 5.1.** [34] *A graph is chordal if and only if it has a perfect elimination ordering.*

**Theorem 5.2.** *For a graph $G$ on $n$ vertices, the following conditions are equivalent:*
*1. $G$ has a perfect elimination ordering.*
*2. $G$ is chordal.*

*3. If $H$ is any induced subgraph of $G$ and $S$ is a vertex separator of $H$ of minimal size then vertices of $S$ induce a clique.*

Chordal graphs can be recognized in linear time.

**Lemma 5.1.** *A chordal graph has at most $n$ maximal cliques.*

Chordal graphs are a subclass of the well known perfect graphs. Other superclasses of chordal graphs include weakly chordal graphs, odd-hole-free graphs and even-hole-free graphs. In fact, chordal graphs are precisely the graphs that are both odd-hole-free and even-hole-free.

Using results of Dirac [30], Fulkerson and Gross [34], Buneman [17], Gavril [37] and Rose et al. [99], we have:

**Theorem 5.3.** *The following statements are equivalent and characterize chordal graphs.*

*(i) $G$ has a simplicial elimination scheme.*

*(ii) Every minimal separator is a clique.*

*(iii) $G$ admits a maximal clique tree.*

*(iv) $G$ is the intersection graph of subtrees in a tree.*

*(v) Any LexBFS provides a simplicial elimination scheme.*

Given an undirected graph $G = (V, E)$, and two non-adjacent vertices $a$ and $b$, a subset $S \subset V$ is an $a, b$-separator if the removal of $S$ separates $a$ and $b$ in distinct connected components. If no proper subset of $S$ is an $a, b$-separator then $S$ is a minimal $a, b$-separator. A minimal separator is a set of vertices $S$ for which there exist non-adjacent vertices $a$ and $b$ such that $S$ is a minimal $a, b$-separator.

It is known that the minimal separators of a chordal graphs are complete subgraphs [30]. Let $G = (V, E)$ be a chordal graph. The clique-graph of $G$, denoted by $C(G) = (V_c, E_c, \mu)$ with $\mu : E_c \to N$ is defined by

(a) The vertex set $V_c$ is the set of maximal cliques of $G$;

(b) The edge $(C_1, C_2)$ belongs to $E_c$ if and only if the intersection $C_1 \cap C_2$ is a minimal $a, b$-separator for each $a \in (C_1 - C_2)$ and each $b \in (C_2 - C_1)$;

(c) The edges of $(C_i, C_j) \in V_c$ are weighted by the cardinality of the corresponding minimal separator $S_{ij}$: $\mu(C_i, C_j) = |S_{ij}|$.

Let $S_{ij} = C_i \cap C_j$ if and only if $S_{ij}$ is a minimal $a, b$-separator for each $a \in C_i - C_j$ and each $b \in C_j - C_i$.

**Lemma 5.2.** *Let $(C_1, C_2, C_3)$ be a 3-cycle in $C(G)$ and let $S_{12}$, $S_{13}$, $S_{23}$ be the associated minimal separators of $G$. Then two of these three minimal separators are equal and included in the third.*

But, the converse of this result is not true.

**Lemma 5.3.** [36] *Let $C(G)$ be the clique graph of the chordal graph $G$. Let $C_1$, $C_2$, $C_3$ be three maximal cliques such that $(C_1, C_2) \in E_c$ and $(C_1, C_3) \in E_c$, then $S_{12} \subset S_{13} \Rightarrow (C_2, C_3) \in E_c$.*

A reversible elimination scheme is just an ordering of the vertices which is simplicial in both directions. A vertex is said to be bisimplicial if its neighbourhood can be partitioned into two cliques. Furthermore, if a graph $G$ admits such a reversible elimination scheme, this implies that each vertex is either simplicial or bisimplicial. Therefore, such a graph cannot contain any claw $(K_{1,3})$ as subgraph.

**Theorem 5.4.** *A graph $G$ admits a reversible ordering if and only if $G$ is proper interval graph.*

### 5.1. Weakly chordal graphs

A graph $G = (V, E)$ is *chordal* if and only if every cycle of length $> 3$ in $G$ has a chord. A graph $G = (V, E)$ is *weakly chordal* if and only if every cycle of length $> 4$ in $G$ and the complement graph $\overline{G}$ has a chord.

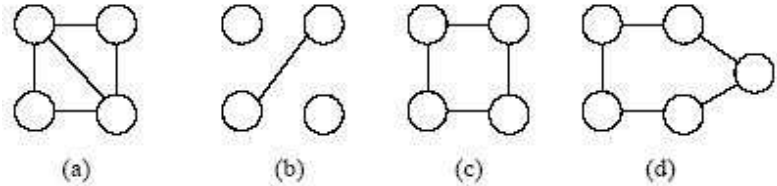

**Figure 11:** (b) is complement of (a). (a) and (b) both are chordal and weakly chordal. (c) is weakly chordal but not chordal. (d) is neither chordal nor weakly chordal.

Weakly chordal graphs were introduced by Hayward [39] in 1985. Both, chordal and weakly chordal graphs are perfect.

**Theorem 5.5.** [39] *Every chordal graph is weakly chordal.*

**Theorem 5.6.** [30] *Any non-complete chordal graph has at least two non-adjacent simplicial vertices.*

It may be verified that all induced subgraphs of a chordal graph also are chordal.

A vertex is LB-simplicial if all the minimal separators included in its neighborhood are cliques [2].

**Theorem 5.7.** [2] *A graph $G = (V, E)$ is weakly chordal if and only if every edge of $E$ is LB-simplicial.*

**Theorem 5.8.** [107] *Weakly chordal graph can be recognized in $O(m^2)$ time, where $m$ is the number of edges of the graph.*

## 6. Permutation graphs

An undirected graph $G = (V, E)$ with vertices $V = \{1, 2, \ldots, n\}$ is called a permutation graph if there exists a permutation $\pi$ on $N = \{1, 2, \ldots, n\}$ such that for all $i, j \in N$,

$$(i - j)(\pi^{-1}(i) - \pi^{-1}(j)) < 0$$

if and only if $i$ and $j$ are joined by an edge in $G$ [42]. Geometrically, the integers $1, 2, \ldots, n$ are drawn in order on a real line called as *upper line* and $\pi(1), \pi(2), \ldots, \pi(n)$ on a line parallel to this line called as *lower line* such that for each $i \in N$, $i$ is directly below $\pi(i)$. Next, for each $i \in V$, a line segment is drawn from $i$ on the lower line to $i$ on the upper line and it is denoted by $l(i)$. Then from definition it follows that there is an edge $(i, j)$ in $G$ if and only if the line segment $l(i)$ for $i$ intersects the line segment $l(j)$ for $j$.

As an illustration, a permutation graph with its permutation representation is considered in Figure 12.

A large amount of works on permutation graphs are done by several scholars [66, 69, 70, 71, 87, 86, 100, 114].

Many important and useful characterizations are available on permutation graphs, some of them are presented below.

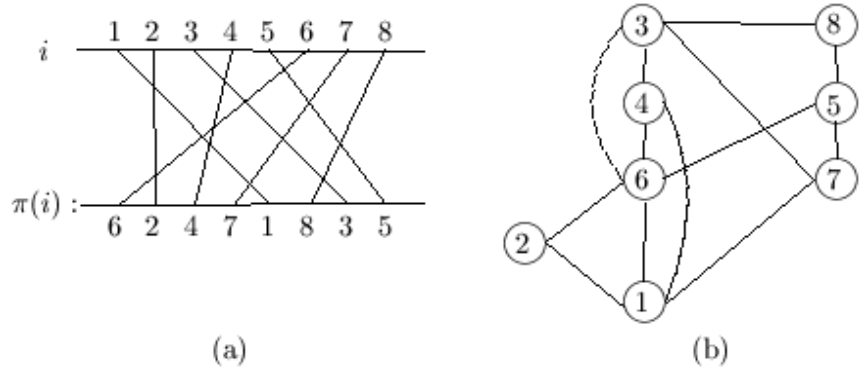

**Figure 12:** (a) A permutation $\pi$, (b) The corresponding permutation graph.

Let $\pi$ be the permutation on $N$ corresponding to the permutation graph $G[\pi]$. If we reverse the sequence $\pi$ then we obtain a graph which is also a permutation graph and it is nothing but the complement of the graph $G[\pi]$. In other words, if $\pi'$ be the reverse order of $\pi$ then $G[\pi'] = G'[\pi]$ [42].

Thus we have the following result.

**Lemma 6.1.** *The complement of a permutation graph is also a permutation graph.*

Another property of the graph $G[\pi]$ is that it is transitively orientable.

An important result proposed by Pnueli [90] is stated below:

**Theorem 6.1.** *An undirected graph $G$ is a permutation graph if and only if $G$ and its complement graph $G'$ are comparability graph.*

One observation between the cliques, stable sets and subsequences of $\pi$ is given below:

**Lemma 6.2.** *The decreasing subsequences of $\pi$ and the cliques of $G[\pi]$ are in one-to-one correspondence. The increasing subsequence of $\pi$ and the stable sets of $G[\pi]$ are in one-to-one correspondence.*

Let $P^2$ be a two dimensional plane whose axes are indexed by $i$ and $\pi^{-1}(i)$. For a given permutation graph $G = (V, E)$, corresponding to each vertex $i \in V = \{1, 2, \ldots, n\}$ we define a point $p_{ij}$ in $P^2$ as $(i, j), j = \pi^{-1}(i)$. The origin of the plane $P^2$ is $p_{00} = (0, 0)$. Let $S(G) = \{p_{ij} : j = \pi^{-1}(i), i = 1, 2, \ldots, n\}$ be the set of all such points in the plane $P^2$. Thus a one to one correspondence between the set of vertices $V$ of the permutation graph $G$ and the set of points $S(G)$ of the point representation of $G$ is established. The point representation of the graph of Figure 12 is shown in Figure 13.

Two points $p_{ij}$ and $p_{kl}$ ( $j = \pi^{-1}(i), l = \pi^{-1}(k)$ ) of $S(G)$ are said to be *non-connected* if either $i < k$ and $j < l$ or $i > k$ and $j > l$. Further $p_{ij}$ is said to be *directly non-connected* with $p_{kl}$ if $p_{ij}$ is non-connected with $p_{kl}$ and there exists no other point $p_{rs} \in S(G)$ $i < r < n, j < s < l$ such that $p_{ij}$ is non-connected with $p_{rs}$ and $p_{rs}$ is non-connected with $p_{kl}$. For example, the points $p_{22}$ and $p_{74}$ of Figure 13 are non-connected as $2 < 7$, $2 < 4$. The points $p_{15}$ and $p_{22}$ are not non-connected as $1 < 2$ but $5 > 2$. The points $p_{22}$ and $p_{43}$ are directly non-connected, but, $p_{22}$ and $p_{74}$ are not directly non-connected as there is a point $p_{43}$ between $p_{22}$ and $p_{74}$ satisfying $2 < 4 < 7$ and $2 < 3 < 4$. Two points are *connected* if they are not non-connected.

**Lemma 6.3.** *Two points $p_{ij}$ and $p_{rs} \in S(G)$ are connected if*
    (a) $i < r$ and $j > s$ or (b) $i > r$ and $j < s$.

A relation between connected points of $S(G)$ and corresponding edges of $G$ is established in the following lemma.

**Lemma 6.4.** *Let $p_{ij}, p_{kl} \in S(G)$ be two points corresponding to two vertices $i, k \in V$. The vertices $i$ and $k$ are connected by an edge in $G$ if and only if $p_{ij}$ and $p_{kl}$ are connected.*

A sequence of points in $S(G)$ is called a *chain* if any point of it is directly non-connected with the next one. A chain of $S(G)$ is said to be *maximal* if it is not contained in any other chain of $S(G)$. A chain of $S(G)$ is said to be *maximum* if its length (the number of points in the chain) is maximum. Let $w(i), i = 1, 2, \ldots, n$ be the weight of the vertex $i$ and $w(p_{ij}) = w(i), i = 1, 2, \ldots, n$ be the weight of the point $p_{ij}$. A chain of $S(G)$ is said to be of *maximum weight* if the sum of the weights of the points of that chain is maximum. A *path* of a tree is an alternating sequence of nodes and edges such that starting and ending nodes are different.

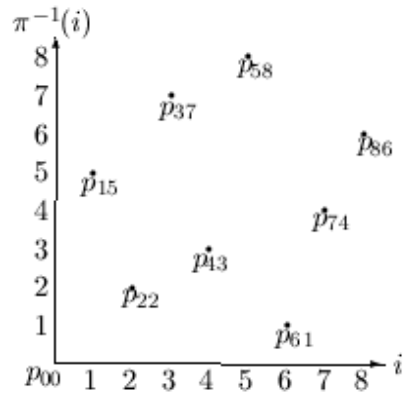

**Figure 13:** Point representation of the permutation graph of Figure 12 in plane $P^2$.

A tree $T(G)$ is constructed recursively as follows:

Let $p_{00}$ be the root of $T(G)$. All the points directly non-connected with $p_{00}$ are taken as the children of $p_{00}$. Let $D_i(G)$ be the set of all directly non-connected points of $p_{ij}$. If $p_{ij}$ is a node of $T(G)$ then its children are the members of $D_i(G)$.

It may be noted that the nodes of $T(G)$ are the members of $\{p_{00}\} \cup S(G)$, but, the number of nodes is not equal to the cardinality of $\{p_{00}\} \cup S(G)$. From the definition of $T(G)$, it is obvious that, if $p_{ij}$ is the parent of $p_{kl}$ then $p_{ij}$ is directly non-connected with $p_{kl}$. One node may appear more than once in the tree $T(G)$. But, no vertices appear more than once in a path starting from a leaf to the root of $T(G)$. Thus the number of nodes of any path starting from a leaf to the root of $T(G)$ is not more than $n$.

From this analogy it is obvious that if the permutation graph $G$ has $n$ maximal independent sets then the total number of nodes of the tree $T(G)$ is $O(n^2)$. This situation occurs for most of the permutation graphs. But, a permutation graph may have more than $O(n)$ maximal independent sets. For example, let $G_1 = (V_1, E_1)$ where $V_1 = \{1, 2, \ldots, n\}$ and $E_1 = \{(i, i+1) : i = 1, 2, \ldots, n-1\}$ be the sets of vertices and edges of a permutation graph. This graph has more than $O(n)$ maximal independent sets. But, this graph is a chain, a tree, an interval graph, a trapezoid graph, etc. In general, this graph may be regarded as a general graph. Let the total numbers of nodes of $T(G)$ be $N$.

The tree $T(G)$ for the graph of Figure 12 is shown in Figure 14.

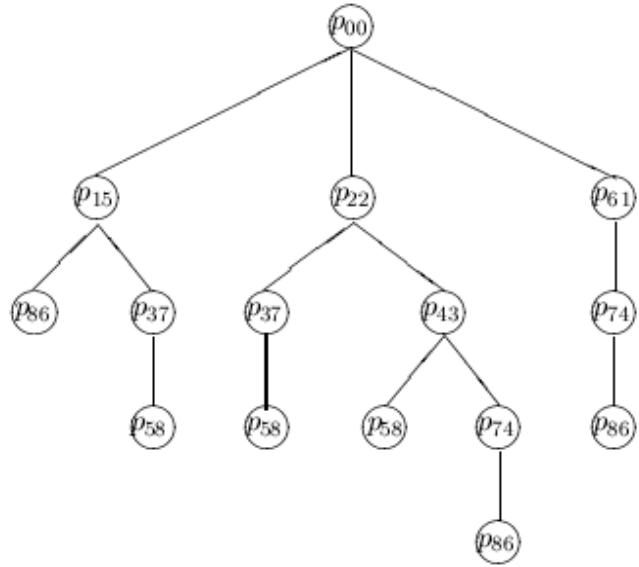

**Figure 14:** The tree $T(G)$ for the graph of Figure 12.

From the definition of $T(G)$ it follows that if any two nodes lie on the same path in the tree $T(G)$ then their corresponding vertices are non-connected in the graph $G$. So the points of any path from a leaf to the root of $T(G)$ form a chain.

**Lemma 6.5.** [104] *The vertices corresponding to the points of any path from a leaf to the root of $T(G)$ form a maximal independent set in $G$.*

**Theorem 6.2.** [82,104] *A maximum weight independent set of a permutation graph with $n$ vertices can be computed in $O(n^2)$ time, provided the graph has at most $O(n)$ maximal independent sets.*

For a weighted permutation graph $G$ the maximum weight $k$-independent set problem is defined as follows:

Given a weighted permutation graph $G$, such that each vertex $v \in V$ has a positive weight $wt(v)$, the maximum weight $k$-independent (MWkI) set problem on $G$ is to find a set of $k$ disjoint partitions $H_1, H_2, \ldots, H_k$ of $G$ such that its weight $\sum_{v \in Q_k} wt(v)$ is maximum, where $Q_k = \cup_{i=1}^{k} H_i$.

**Theorem 6.3.** *The maximum weight $k$-independent set problem can be solved in $O(kn^2)$ time on permutation graph [100] and $O(kn\sqrt{\log c} + m)$ time on interval graph, where m and c represent respectively the number of edges and weight of the longest path of the [79,81].*

The depth-first search (DFS) is an important problem in graph theory. A very efficient algorithm is designed in [66] for DFS on permutation graphs, which also constucts a depth-first tree.

**Theorem 6.4.** [66] *The depth-first tree of a permutation graph can be constructed in* $O(n)$ *time.*

For a given subset $T$ of $V$, called as set of target vertices, a set $S \subset V$ is said to be a *Steiner set* for $T$ in $G$ if

(i) $S$ is a subset of $V - T$, i.e. $S \subseteq V - T$,

(ii) the subgraph induced by $S \cup T$ in $G$ is connected.

The Steiner set $S$ is said to be *'minimum cardinality Steiner set'*, if the cardinality of $S$ is minimum. A spanning tree of a connected subgraph induced by $S \cup T$ in $G$ is called a *Steiner tree*.

The minimum cardinality Steiner set problem is the problem of finding the minimum number of vertices to connect a given set of target vertices $T$. Finding minimum Steiner set of an arbitrary graph is known to be NP-complete.

**Theorem 6.5.** [69] *The minimum cardinality Steiner tree and Steiner set can be computed in* $O(n)$ *time.*

**Theorem 6.6.** [93] *Minimum cardinality 2-neighbourhood covering set of a permutation graph can be computed in* $O(n + \overline{m})$*, where* $\overline{m}$ *is the number of edges of the complement graph.*

Let $k > 1$ be a positive integer which is given together with the input graph. For two distinct vertices $u$ and $v$ in $G$, the distance $d(u,v)$ between $u$ and $v$ is the length (i.e. number of edges) of a shortest path between $u$ and $v$. A vertex $u \in V$ is said to be $k$ *-dominates* the vertex $v \in V$ if $d(u,v) \le k$. A subset $D \subseteq V$ is a $k$ *-dominating set* in G if for every $v \in V$ there is at least one $u \in D$ with $d(u,v) \le k$. $D$ is a *total* $k$ *-dominating set* of $G$ if every vertex $v \in V$ is $k$ -dominated by a vertex in $D$ and for $u \in D$ there is at least one vertex $w(\ne u) \in D$ with $d(u,w) \le k$. Note that the subgraph induced by a total $k$ -dominating set may have isolated vertices for $k > 1$.

**Theorem 6.7.** [95] *The minimum cardinality* $k$ *-domination set and a minimum cardinality total* $k$ *-domination set can be computed in* $O(n + \overline{m})$ *time.*

A subset $D$ of $V$ is called a distance $k$ -dominating set of $G$ if each $x \in V \setminus D$ is within distance $k$ from some vertices of $D$. The minimum cardinality of a distance $k$ -dominating set in $G$ is the distance $k$ -domination number of $G$, denoted by $\gamma_k(G)$. The distance $k$ *-domination problem* is to find a $\gamma_k(G)$ in $G$. The distance 1-domination number $\gamma_1(G)$ is the usual domination number $\gamma(G)$. In general, determining $\gamma_k(G)$ is NP-complete [35].

**Theorem 6.8.** [95] *The minimum cardinality distance* $k$ *-domination set on permutation*

73

*graph can be determined in* $O(n^2)$ *time.*

Like interval graph, the all-pair shortest path problem on permutation graph can also be solved in optimal time as stated below.

**Theorem 6.9.** [67] *The all-pair shortest paths problem can be solved on permutation graphs in* $O(n^2)$ *time. Also, the average distance can be determined in* $O(n^2)$ *time.*

**Theorem 6.10.** [7] *The time complexity to find the next-to-shortest path between any two specified vertices in permutation graph is* $O(n^2)$.

For a fixed positive integer $k$, a $k$-tuple domination set of a graph $G = (V, E)$ is a subset $D \subseteq V$ such that every vertex in $V$ is dominated by at least $k$ vertices in $D$. The $k$-tuple domination number $\gamma_{\times k}(G)$ is the minimum size of a $k$-tuple dominating set of $G$. The special case when $k = 1$ is the usual domination. The case when $k = 2$ was called double domination or 2-tuple domination. A 2-tuple dominating set $D_2$ is said to be minimal if there does not exist any $D' \subset D_2$ such that $D'$ is a 2-tuple dominating set of $G$. A 2-tuple domination set $D_2$, denoted by $\gamma_{\times k}(G)$, is said to be minimum, if it is minimal as well as it gives 2-tuple domination number. For any graph $|D_2| \geq 2$ for $n \geq 2$.

**Theorem 6.11.** [11] *The minimum 2-tuple domination set of a permutation graph can be computed in* $O(n^2)$ *time.*

## 7. Trapezoid graphs

A trapezoid $T_i$ is defined by four corner points $[a_i, b_i, c_i, d_i]$, where $a_i < b_i$ and $c_i < d_i$ with $a_i, b_i$ lying on top line and $c_i, d_i$ lying on bottom line of a rectangular channel. An undirected graph $G = (V, E)$ with vertex set $V = \{v_1, v_2, \ldots, v_n\}$ and edge set $E = \{e_1, e_2, \ldots, e_m\}$ is called a trapezoid graph if a trapezoid representation can be obtained such that each vertex $v_i$ in $V$ corresponds to a trapezoid $T_i$ and $(v_i, v_j) \in E$ if and only if the trapezoids $T_i$ and $T_j$ corresponding to the vertices $v_i$ and $v_j$ intersect. For simplicity the vertices $v_1, v_2, \ldots, v_n$ are represented respectively by 1, 2, …, $n$. Thus the edge $(i, j) \in E$ if and only if $T_i$ and $T_j$ intersect in the trapezoid representation. Figure 15 illustrates a trapezoid graph and its trapezoid representation consisting of seven trapezoids $T_1, T_2, \ldots, T_7$. It is interesting to note that if $a_i = b_i$ and $c_i = d_i$ then the corresponding trapezoid $T_i$ reduces to a straight line. So, in this way, if all the trapezoids reduce to straight lines the corresponding trapezoid graph reduces to nothing but a permutation graph. For simplicity, we assume that the corner points on the trapezoid representation are all distinct and so they can be given consecutive positions $1, 2, \ldots, 2n$ from left to right on both channels. In addition to this we may label these $n$ trapezoids in increasing order of their right corner points on top channel, i.e. for two trapezoids $T_i$ and $T_j, i < j$ if and only if $b_i$ lies on the left of $b_j$.

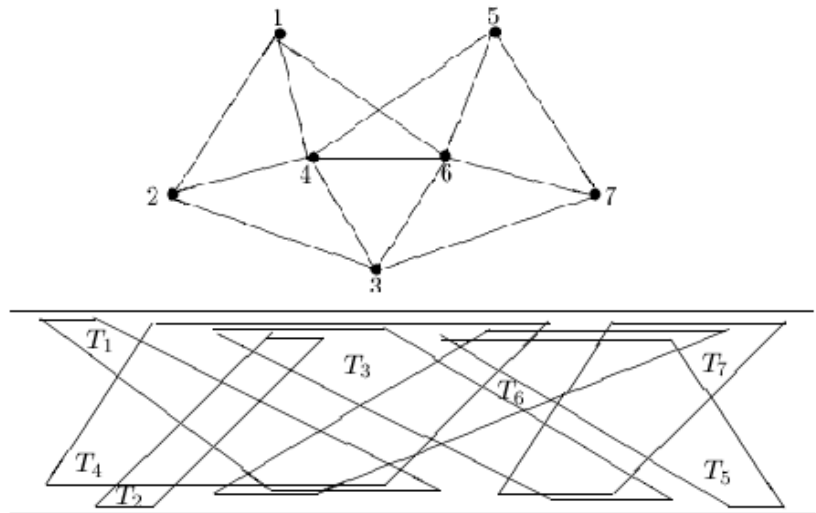

**Figure 15:** A trapezoid graph $G$ and its trapezoid representation

Trapezoid graphs can be recognized in $O(n^2)$ time by Ma and Spinrad's algorithm [59].

As shown in [28], trapezoid graphs can be used for modelling a *channel routing* problem in VLSI, in a single-layer-per-net model. A *channel* consists of a pair of horizontal lines with points or terminals on each line numbered from 1 to $n$. All the terminals with the same label constitute a *net*. A *routing* is a connection of every net by wires inside the channel such that no two wires from different nets overlap (see Figure 16). A routing is allowed to use more than one layer. This channel routing problem is equivalent to the minimum colouring problem of a trapezoid graph, where each net is represented by a trapezoid. The *single module $k$-planer (i.e. $k$ layers) subset* problem in VLSI is to assign maximum possible nets in $k$ layers inside a channel in such a way that no two nets, assigned in any of the $k$ layers overlap each other where the routing region is either a channel or bounded by a straight line and a solid module.

A trapezoid graph with $n$ vertices can be represented geometrically either by,

1. a set of $n$ trapezoids drawn inside a rectangular channel or by,
2. a set of $n$ segments drawn on a two dimensional plane or by,
3. a set of $n$ boxes drawn on a two dimensional plane or by,
4. a permutation diagram $\pi$ of $2n$ lines drawn inside a channel.

### 7.1. Trapezoid representation

Let $T = \{T_1, T_2, \ldots, T_n\}$, be the set of $n$ trapezoids where trapezoid $T_i$ is represented in the trapezoid representation by four corner points $[a_i, b_i; c_i, d_i]$, $a_i, b_i (a_i < b_i)$ lying on the top line and $c_i, d_i (c_i < d_i)$ lying on the bottom line of a rectangular channel (see

Figure 15). Without any loss of generality we assume the following:

1. a trapezoid contains four distinct corner points and that no two trapezoids share a common end point,

2. trapezoids in the trapezoid representation and vertices in the trapezoid graph are one and same thing,

3. the trapezoids in the trapezoid representation $T$ are indexed by increasing right end points on the top line, i.e. for any two trapezoids $T_i$ and $T_j$ in the trapezoid representation $i < j$ if and only if $b_i < b_j$.

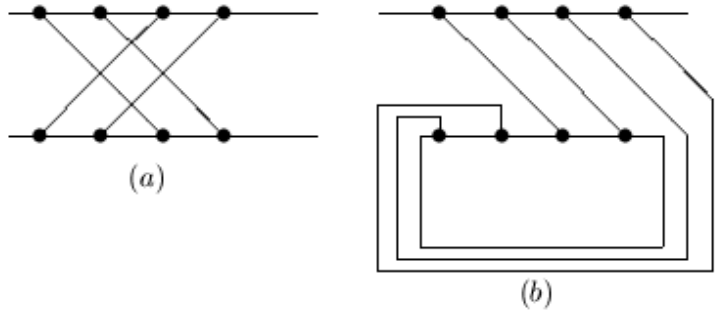

**Figure 16:** (a) A routing instance, where the routing region is a channel. (b) A routing instance, where the routing region is bounded by a straight line and a module.

This kind of ordering gives the following result which is quite useful in designing efficient algorithms.

*In a trapezoid graph $G$ if any three vertices $i, j, k$ are such that $i < j < k$ and $(i, k) \in E$ then either $(i, j) \in E$ or $(j, k) \in E$.*

This ordering is sometimes called as *cocomparability ordering*. It is shown in [108] that in a cocomparability graph this ordering can be implemented in an $O(n^2)$ time. But for a trapezoid graph, this ordering can be implemented in only $O(n)$ time with the help of its trapezoid representation.

The adjacency relation between any two vertices can be tested using the following result:

*Let $i$ and $j$ be two vertices of a trapezoid graph $G$. Then two vertices $i$ and $j$ are not adjacent if and only if either*

(i) $b_i < a_j$ and $d_i < c_j$ or (ii) $b_j < a_i$ and $d_j < c_i$.

Otherwise the vertices $i$ and $j$ are adjacent.

Therefore, instead of storing a trapezoid graph, using adjacency matrix or adjacency list, one can store the trapezoid representation of the trapezoid graph using only $4n$ units of memory. The adjacency relation can be tested in $O(1)$ time.

### 7.2. Segment representation

Segment representation of a trapezoid graph is nothing but a transformation of its trapezoid representation. In segment representation, a trapezoid $T_i[a_i, b_i; c_i, d_i]$ is transformed to a unique line segment $p_i q_i$ where $p_i, q_i$ are two points on a two dimensional plane having coordinates $(a_i, c_i), (b_i, d_i)$ respectively. Segment representation of the graph of Figure 15 is shown in Figure 17.

In segment representation, the segments $p_i q_i$ and $p_j q_j$ are called *disjoint* if and only if either, $b_i < a_j$ and $d_i < c_j$ or, $b_j < a_i$ and $d_j < c_i$. Therefore, in a segment representation if the origin be shifted to any point $q_i, q_i \in S$ then all line segments $p_j q_j$ lying entirely on the first quadrant are disjoint from $p_i q_i$. On the other hand if the origin be shifted to any point $p_i$, $p_i \in S$ then all line segments $p_j q_j$ lying entirely on the third quadrant are also disjoint from $p_i q_i$. Any other line segments are called *joint* to the segment $p_i q_i$.

Therefore, in segment representation, adjacency relation between vertices $i$ and $j$ can be tested using the following result:

In a trapezoid graph $G$, for any two vertices $i$ and $j$, $(i, j) \notin E$ if and only if $p_i q_i$ is not adjacent to $p_j q_j$ in corresponding segment representation. Otherwise $(i, j) \in E$.

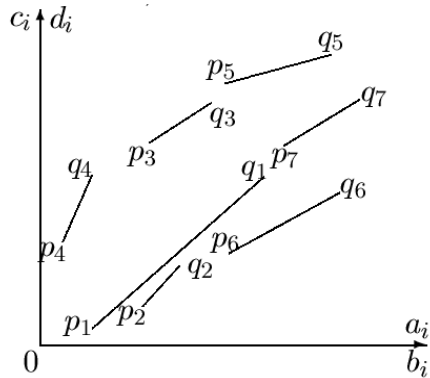

**Figure 17:** The segment representation of the graph of Figure 15

Hence in a segment representation, adjacency relation can be tested in $O(1)$ time.

### 7.3. Box representation

If $x = (x_1, x_2)$ and $y = (y_1, y_2)$ are points in $R^2$, then $x$ is said to be *dominated* by $y$, denoted as $x < y$, if $x_i$ is less than $y_i$ for $i = 1, 2$. The order thus given between points in $R^2$ is called *dominance order*. This order can be extended to *boxes*, i.e. sets of the form $\{(x_1, x_2) \in R^2 : l_1 \le x_1 \le u_1, l_2 \le x_2 \le u_2\}$ where $(l_1, l_2)$ is the *lower corner* and $(u_1, u_2)$ is the *upper corner* of the box.

A box $b$ dominates another box $b'$ if the lower corner of $b$ dominates the upper corner of $b'$. Note that points may be understood as boxes where the lower and upper corner coincide. If one of the two boxes dominates the other we say that they are *comparable*. Otherwise they are *incomparable*. Now the vertices of a trapezoid graph may be represented by boxes with two boxes incomparable if and only if the corresponding vertices are joined by an edge.

In Figure 18, the trapezoid graph of Figure 15 is illustrated with a dominance order $P_G$ and its trapezoid representation, box representation are also given in the same figure. It is easy to observe that with the use of dominance order the vertices 1,2,3,4,5,6,7 in Figure 15 are renamed as vertices 1,4,2,6,3,7,5 respectively in Figure 18.

What makes the box representation useful is the additional dominance order on boxes that may be exploited by sweep line algorithms, where all computations are done in a single sweep.

## 7.4. Permutation representation

From trapezoid representation of a trapezoid graph a permutation diagram can be obtained with the use of the concept of *vertex splitting*. In a trapezoid graph, the vertex splitting process replaces each vertex $v$ by two new vertices $v_1$ and $v_2$ where in the trapezoid representation the trapezoid representing $v$ is replaced by two lines representing $v_1$ and $v_2$ respectively. Thus it may be seen that the trapezoid representation is evolving into a permutation graph representation.

The concept of vertex splitting was first introduced by Cheah and Corneil [23]. They also proved that a graph is a trapezoid graph if and only if after an appropriate sequence of vertex splitting a permutation graph is obtained with a specific condition. Thus with the use of the concept of vertex splitting a trapezoid representation of $n$ trapezoids will be transformed to a permutation diagram $\pi$ of $2n$ lines.

**Theorem 7.1.** [26] *Trapezoid graphs are weakly chordal.*

**Theorem 7.2.** [53] *All cut vertices of a trapezoid graph can be computed correctly in $O(n)$ time.*

The very common problem in graph theory is all pairs shortest path problem. This problem has been solve for trapezoid graph in [67]. The time complexity needed to find all pairs shortest distances is stated below.

**Theorem 7.3.** [67] *The time complexity to find all pairs shortest distances on trapezoid graphs is $O(n^2)$.*

**Theorem 7.4.** [6] *The time complexity to find the next-to-shortest path between any two vertices $u$ and $v$ in trapezoid graph is $O(n^2)$.*

**Theorem 7.5.** [13] *All maximal cliques of a trapezoid graph can be generated in*

$O(n^2 + \gamma n)$ time, where $n$ is the number of vertices of the graph and $\gamma$ is the output size.

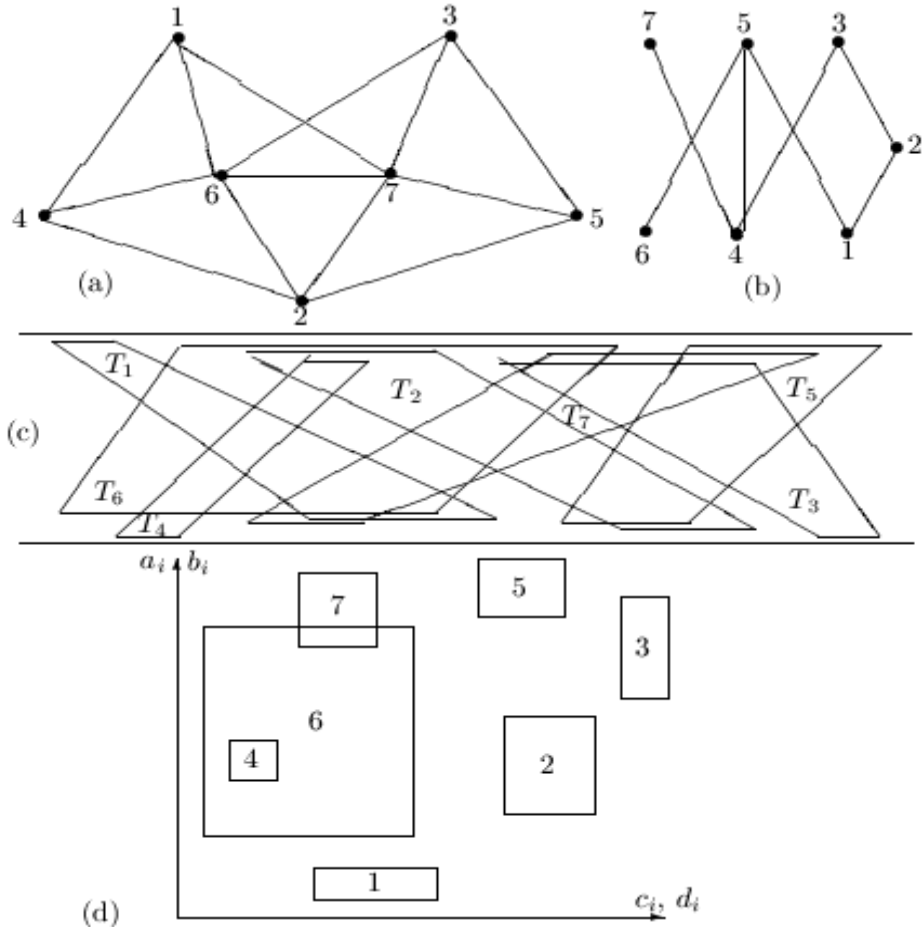

**Figure 18:** (a)The trapezoid graph of Figure 15 when vertices are given the dominance order $P_G$ (b) The dominance order $P_G$ (c) Its trapezoid representation and (d) Its box representation.

**Theorem 7.6.** [51] *All maximal independent sets of a trapezoid graph $G$ with $n$ vertices can be computed in $O(n\overline{m} + \alpha n)$ time, where $\overline{m}$ and $\alpha$ denote respectively the total number of edges in $\overline{G}$, the complement graph of the graph $G$, and the number of maximal independent sets in $G$.*

**Theorem 7.7.** [52] *The maximum weight $k$-independent set problem can be solved for trapezoid graph in $O(kn^2)$ time. In particular, maximum weight 2-independent set* *problem can be solved in* $O(n^2)$ *time.*

A *clique* of an undirected graph $G = (V, E)$ is a complete subgraph of $G$, and a *clique cover* of $G$ is a partition of $V$ such that each set in the partition is a clique. A clique cover with the minimum cardinality $k (k \leq n)$ is known as a *minimum clique cover* (MCC). This number $k$ is called the *clique cover number.*

The minimum clique cover is a well known NP-complete problem on general graphs [35, 42]. However, it can be solved in polynomial time for some special classes of graphs, like chordal graphs, interval graphs, circular-arc-graphs, circular permutation graph, etc.

**Theorem 7.8.** [54] *A minimum clique cover of the trapezoid graph* $G$ *can be computed in* $O(n^2)$ *time.*

**Theorem 7.9.** [54] *The diameter and center of a trapezoid graph* $G = (V, E)$ *can be computed in* $O(dn)$ *time where* $d$ *is the degree of vertex 1.*

It is shown in [25] that every connected AT-free graph contains a dominating pair and the vertices which achieve the diameter are said to form a dominating pair. As trapezoid graphs belong to a subclass of AT-free graphs so the following results.

**Theorem 7.10.** [54] *Every trapezoid graph has at least one dominating pair.*

**Theorem 7.11.** [54] *The vertices* $u, v$ *of a trapezoid graph form a dominating pair if* $\delta(u, v) = diam(G)$.

A *diameter path* in a graph is a shortest path whose length is equal to the diameter of the graph.

**Theorem 7.12.** [54] *If* $p(u, v)$ *is a diameter path in a trapezoid graph* $G$, *then* $p(u, v)$ *is minimum connected dominating path.*

**Theorem 7.13.** [54] *Dominating pairs and minimum connected dominating paths of a trapezoid graph can be computed in* $O(dn)$ *time.*

**Theorem 7.14.** [14] *The set of all hinge vertices of a trapezoid graph with* $n$ *vertices can be computed in* $O(n \log n)$ *time using* $O(n)$ *space.*

**Theorem 7.15.** [15] *A spanning tree of a trapezoid graph can be computed in* $O(n)$ *time.*

**Theorem 7.16.** [9] *The time complexity to find tree 4-spanner on trapezoid graphs is* $O(n)$.

**Theorem 7.17.** [10] *The time complexity to find tree 3-spanner on trapezoid graphs is* $O(n^2)$.

**Theorem 7.18.** [96] *The minimum cardinality conditional covering set on trapezoid graphs can be determined in* $O(n^2)$ *time.*

## 8. Unit disk graphs

Unit disk graph (UDG) is a popular model that enabled the development of efficient algorithms for crucial networking problems. A unit disk is a closed disk of radius 1 in the plane. Unit disk graphs can be represented in two ways:

(a) in the intersection model the vertices are unit disks in the plane and two of them are adjacent if and only if the disks intersect, and

(b) in the distance model, the vertex set is a point set in the plane, and any two vertices are adjacent if and only if their distance is at most 1.

A unit disk graph as a concrete geometric object, that is, the vertex set is a subset of points in two-dimension plane. So, every vertex is a point in the plane.

There is another variation of UDG called $r$-UDG. In a $r$-UDG, two nodes are connected if and only if their distance is at most $r$, for some $r > 0$.

UDG is known to be unrealistic in the context of physical constraints created by interferences in wireless ad-hoc networks. The interferences created by simultaneous radio transmissions may prevent nodes that belong to the $r$-neighborhood of a transmitter to hear its message. The most realistic model capturing the physical interference constraints on the other hand is known as Signal Interference plus Noise Ratio (SINR in short). In SINR, each node has some given emission power level. This model assumes that the signal transmitted decays as a power greater than 2 of the distance. In this model, a node $u$ receives the signal of a node $v$ if and only if the ratio of the signal of $u$ over the noise to which $v$ is subject to above some given threshold.

A very less work has been done on UDG. Some of them are discussed below.

The upper bound of chromatic number of unit disk graphs is given in terms of the clique number. The best known bound is:

**Theorem 8.1.** [89] *A unit disk graph* $G$ *can be coloured with at most* $3\omega(G) - 2$ *colours, where* $\omega(G)$ *is the size of largest clique.*

Another result is also available in terms of stability number as stated below:

**Theorem 8.2** *A unit disk graph* $G$ *with* $\alpha(G) \le 2$ *can be coloured with at most* $\frac{3}{2}(G)$ *colours, where* $\alpha(G)$ *represents the stability number.*

The set of vertices of a UDG can be discomposed into three non-disjoint cliques.

**Lemma 8.1.** *Let* $G$ *be a unit disk graph with* $\alpha(G) \le 2$. *Then* $V(G)$ *is the union of three cliques, two of which contain a common vertex.*

## 9. Graph with boxicity $k$

An interval on the real line can be generalized to a $k$-box in $R^k$. A $k$-box $B = (R_1, R_2, \ldots, R_k)$, where each $R_i$ is a closed interval on the real line, is defined to be the Cartesian product $R_1 \times R_2 \times \cdots \times R_k$. If each $R_i$ is a unit length interval, then the graph $B$ is called $k$-cube. Thus, 1-boxes are just closed intervals on the real line whereas 2-boxes are axis-parallel rectangles in the plane. The parameter boxicity of a graph $G$, denoted as $box(G)$, is the minimum integer $k$ such that $G$ is an intersection graph of $k$-boxes. Similarly, the cubicity of $G$, denoted as $cub(G)$, is the minimum integer $k$ such that $G$ is an intersection graph of $k$-cubes. Thus, interval graphs are the graphs with boxicity at most 1 and unit interval graphs are the graphs with cubicity at most 1. These parameters were introduced by Roberts in 1969 [97].

Any graph can be represented as the intersection graph of a set of rectangles in $d$ dimensional plane where $d \geq 1$. Boxicity-2 graphs are called the rectangle intersection graphs.

1. A graph has boxicity one if and only if it is an interval graph.
2. Every outerplanar graph has boxicity at most two .
3. Every planar graph has boxicity at most three .
4. If a bipartite graph has boxicity two, it can be represented as an intersection graph of axis-parallel line segments in the plane.
5. The upper bound of the boxicity $d$ is $d \leq \tau + 1$, where $\tau$ is the tree-width of the graph.

A graph which is not an interval graph can be represented by 2-boxes. For example, the graph $C_4$, can be seen to be an intersection graph of 2-boxes (see Figure 19).

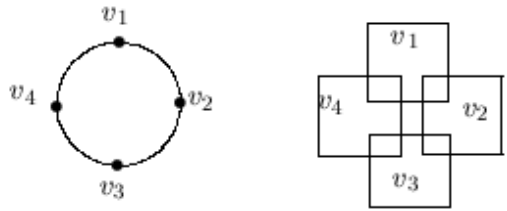

**Figure 19:** A 2-box representation for $C_4$

The following lemma due to Roberts [97] is very interesting.

**Lemma 9.1.** [97] *For any graph $G$, $box(G) \leq k$ if and only if there exists $k$ interval graphs $I_1, \ldots, I_k$ such that $G = I_1 \cap I_2 \cap \cdots \cap I_k$.*

Roberts also proved the following theorem.

**Theorem 9.1.** [97] *For any graph $G$ with $n$ vertices $box(G) \leq n$.*

The problem to represent a graph into boxes is NP-complete. But, an efficient solution or approximation algorithm is available for lower values of $k$. Also, the decision version of the problem, i.e. to test whether the boxicity of a given graph is at most some given value $k$, even for $k$=2, is NP-complete. Chandran et al. [20] described algorithms for finding representations of arbitrary graphs as intersection graphs of boxes, with a dimension that is within a logarithmic factor of the maximum degree of the graph.

**Theorem 9.2.** **[21]** *Given a graph $G$ of $n$ vertices with maximum degree $\Delta$,* $box(G) \leq \lceil (\Delta + 2) \ln n \rceil$.

## 10. Other intersection graphs

### 10.1. Circle graphs

The *circle graph* (see Figure 20) is the intersection graph of a set of chords of a circle. For each chord there is a vertex in the graph and any two vertices are connected if and only if corresponding chords intersect.

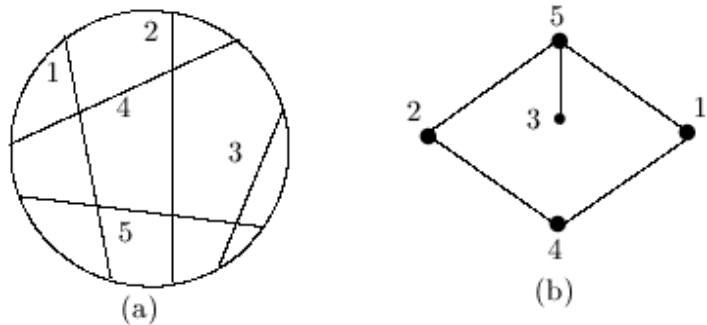

**Figure 20:** (a) A set of chords, and (b) the corresponding circle graphs

In [109], Spinrad presented an $O(n^2)$ time algorithm to test whether a given undirected graph is a circle graph. If the graph is circle then this algorithm constructs a set of chords that represent it.

Tiskin [111] has shown that a maximum clique of a circle graph can be found in $O(n \log^2 n)$ time, while Nash and Gregg [72] have shown that a maximum independent set of an unweighted circle graph can be determined in $O(n \min\{d, \alpha\})$ time, where $d$ is a parameter of the graph known as its density, and $\alpha$ is the independence number of the circle graph.

### 10.2. Line graphs

The *line graph* of a graph $G$ is generally denoted by $L(G)$ that represents the adjacencies between edges of $G$. The line graph is defined as follows:

Given a graph $G$, its line graph $L(G)$ is a graph such that

(i) for each edge of $G$ there is a vertex of $L(G)$; and

(ii) two vertices of $L(G)$ are adjacent if and only if corresponding edges share a common endpoint in $G$.

That is, line graph is the intersection graph of the edges of $G$, representing each edge by the set of its two endpoints.

The line is also known as *theta-obrazom, the covering graph, the derivative, the edge-to-vertex dual, the conjugate, and the representative graph, as well as the edge graph, the interchange graph, the adjoint graph, and the derived graph.*

Let us consider a graph $G$ shown in Figure 21(a). The edges of this graph are $e_1 = (1,2)$, $e_2 = (1,5)$, $e_3 = (1,6)$, $e_4 = (2,3)$, $e_5 = (3,4)$, $e_6 = (4,5)$, $e_7 = (4,6)$. Seven vertices are drawn (see Figure 21(b)) for seven edges. Since $e_1$ and $e_2$ have a common vertex 1, so there is an edge between $e_1$ and $e_2$. Again, there is no common vertex between the edges $e_2$ and $e_5$ so there is no edge between $e_2$ and $e_5$. In this way the entire line graph $L(G)$ (Figure 21(b)) is constructed from $G$.

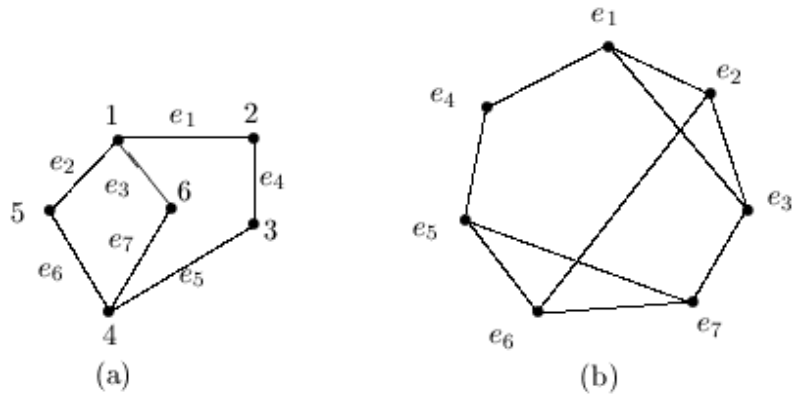

**Figure 21:** (a) A graph $G$, (b) Line graph of $G$

The line graph of the complete graph $K_n$ is the triangular graph. Every line graph is a claw-free graph. Some of the properties of claw-free graphs are generalizations of those of line graphs. The line graph of a bipartite graph is perfect. The line graphs of bipartite graphs form one of the key building blocks of perfect graphs.

Some important results of line graph are presented below:

1. The line graph of a connected graph is connected.

2. A maximum independent set in a line graph corresponds to maximum matching in the original graph.

3. The edge chromatic number of a graph $G$ is equal to the vertex chromatic number of its line graph $L(G)$.

4. The line graph of an edge-transitive graph is vertex-transitive.

5. If a graph G has an Euler cycle, that is, if $G$ is connected and has an even number of edges at each vertex, then the line graph of $G$ is Hamiltonian.

6. The line graphs of trees are exactly the claw-free block graph.

It is very interesting that line graph of line graph is not the original graph, it gives some other graph. $L$ can be treated as an operator. Rooij and Wilf [98] consider the sequence of graphs $G, L(G), L(L(G)), L(L(L(G))), \ldots$
They proved the following behaviours of this sequence.

1. If $G$ is a cycle graph then $L(G)$ and each subsequent graph in this sequence is isomorphic to $G$ itself. These are the only connected graphs for which $L(G)$ is isomorphic to $G$.

2. If $G$ is a claw $K_{1,3}$, then $L(G)$ and all subsequent graphs in the sequence are triangles.

3. If $G$ is a path graph then each subsequent graph in the sequence is a shorter path until eventually the sequence terminates with a graph having only one vertex.

4. In all remaining cases, the sizes of the graphs in this sequence eventually increase.

## 10.3. String graphs

In graph theory, a *string graph* is an intersection graph of curves in the plane, each curve is called a "string". Let $S$ be a set of strings in a plane such that no three strings intersect at a single point. Draw a vertex for each string and an edge for each intersecting pair of strings. This intersection graph is called *string graph*. The mathematical study of string graphs began with work of Ehrlich et al. [32]. The recognition of string graphs is NP-complete [75].

It is easy to verify that every planar graph is a string graph. The string representation is described as follows: draw a string for each vertex that loops around the vertex and around the midpoint of each adjacent edge. For any edge $(u,v)$ of the graph, the strings for $u$ and $v$ cross each other twice near the midpoint of the edge $(u,v)$, and there are no other crossings, so the pairs of strings that cross represent exactly the adjacent pairs of vertices of the original planar graph. Chalopin et al. [22] proved that every planar graph has a string representation in which each pair of strings has at most one crossing point, unlike the representations described above.

Let $P$ be a set of simple paths on a (rectangular) grid. The vertex intersection graph VPG($P$) of $P$ has vertex set $V$, where every vertex $v \in V$ corresponds to a path $P_v \in P$, and edge set E, where $(u,v) \in E$ if and only if the corresponding paths $P_u$ and $P_v$ intersect, i.e. $E = \{(u,v) \mid u, v \in V, u \neq v, P_u \cap P_v \neq \phi\}$. We call a graph G a VPG graph if $G \cong$ VPG($P$), for some $P$. If $P$ is a set of simple paths on a grid, where each path has at most $k$ bends ($90°$ turns), then the graph $G$ is called $B_k - VPG$.

$B_k - VPG$ graphs are related to several other families of intersection graphs that have been studied in the literature. It is rather simple to prove that the string graphs are equivalent to the VPG graphs when there is no restriction on the number of bends per path in the grid. Interval graphs and trees are both subfamilies of $B_0 - VPG$, and the so called grid intersection graphs [48] are equivalent to the bipartite $B_0 - VPG$ graphs.

Circle graphs are a subfamily of string graphs, and it is easy to show that they are contained in the class $B_1 - VPG$. This immediately implies that the coloring problem is NP-complete on $B_1 - VPG$ graphs. We prove the stronger result that the coloring problem is NP-complete for $B_0 - VPG$ graphs. It is proved that every planar graph is a $B_3 - VPG$ graph [1]. Another connection between planar graphs and string graphs began when Scheinerman and West [106] conjectured that planar graphs are contained in the family of segment graphs (SEG), the intersection graphs of straight-line segments in the plane with an arbitrary number of directions. Recently, Chalopin et al. [22] proved Scheinerman's conjecture.

# REFERENCES

1. A.Asinowski, E.Cohen, M.C.Golumbic, V.Limouzy, M.Lipshteyn and M.Stern, Vertex intersection graphs of paths on a grid, *Journal of Graph Algorithms and Applications*, 16 (2) (2012) 129-150.

2. A.Berry, J.Paul Bordat and P.Heggernes, Recognizing weakly triangulated graphs by edge separability, In LNCS 1851, *Proceedings of Seventh Scandinavian Workshop on Algorithm Theory*, 139-149, 2000.

3. Y.Aumann, M.Lewenstein, O.Melamud, R.Y.Pinter and Z.Yakhini, Dotted interval graphs and high throughput genotyping, In 16th annual *ACM-SIAM Symposium on Discrete Algorithms*, 339-348, 2005.

4. Y.Aumann, M.Lewenstein, O.Melamud, R.Y.Pinter and Z.Yakhini, Dotted interval graphs, *ACM Transactions on Algorithms*, 8(2) (2012) Article no. 9.

5. R.Bar-Yehuda, M.M.Halldórsson, J.Naor, H.Shachnai and I.Shapira, Scheduling split intervals, *SIAM J. Comput.*, 36(1) (2006) 1-15.

6. S.C.Barman, M.Pal and S.Mondal, An efficient algorithm to find next-to-shortest path on trapezoidal graph, *Advances in Applied Mathematical Analysis*, 2(2) (2007) 97-107.

7. S.C.Barman, M.Pal and S.Mondal, An efficient algorithm to find next-to-shortest path on permutation graph, *Journal of Applied Mathematics and Computing*, 31 (2009) 369-384.

8. S.C.Barman, M.Pal and S.Mondal, The $k$-neighbourhood-covering problem on interval graphs, *International Journal of Computer Mathematics*, 87(9) (2010) 1918-1935.

9. S.C.Barman, S.Mondal and M.Pal, A linear algorithm to construct a tree 4-spanner on trapezoid graphs, *Intern. J. of Computer Mathematics*, 87 (2010) 743-755.

10. S.C.Barman, S.Mondal and M.Pal, Computation of a tree 3-spanner on trapezoid graphs, *Annals of Pure and Applied Mathematics*, 2(2) (2012) 135-150.

11. S.C.Barman, M.Pal and S.Mondal, Minimum 2-tuple dominating set of permutation graphs, *J. Appl. Math. Comput.,* 43 (2013) 133-150.

12. D.Bera, M.Pal and T.K.Pal, A parallel algorithm for computing all hinge vertices on interval graphs, *Korean J. Computational and Applied Mathematics*, 8(2) (2001) 295-309.

13. D.Bera, M.Pal and T.K.Pal, An efficient algorithm for generating all maximal cliques on trapezoid graphs, *Intern. J. Computer Mathematics*, 79 (10) (2002) 1057-1065.

14. D.Bera, M.Pal and T.K.Pal, An efficient algorithm for finding all hinge vertices on trapezoid graphs, *Theory of Computing Systems*, 36(1) (2003) 17-27.

15. D.Bera, M.Pal and T.K.Pal, An optimal PRAM algorithm for a spanning tree on trapezoid graphs, *J. Applied Mathematics and Computing*, 12(1-2) (2003) 21-29.

16. K.Bogart, M.Jacobson, L.Langley and F.McMorris, Tolerance orders and bipartite unit tolerance graphs, *Discrete Math.*, 226 (2001) 35-50.

17. P.Buneman, A characterization of rigid circuit graphs, *Discrete Math.*, 9 (1974) 205-212.

18. A.Butman, D.Hermelin, M.Lewenstein and D.Rawitz, Optimization problems in multiple-interval graphs, *ACM Trans. Algorithms*, 6(2) 2010, Article no. 40.

19. M.C.Carlisle and E.L.Loyd, On the $k$-coloring of intervals, *LNCS*, 497, ICCI'91, (1991) 90-101.

20. L.S.Chandran and N.Sivadasan, Geometric representation of graphs in low dimension, In Proceedings of the 12th *Annual International Computing and Combinatorics Conference*, Taipei, Taiwan, 398-407, August 2006.

21. L.S.Chandran and N.Sivadasan, Boxicity and treewidth, *J. Combin. Theory Ser. B,* 97(5) (2007) 733-744.

22. J.Chalopin, D.Gonçalves and P.Ochem, Planar graphs are in 1-STRING, Proceedings of the Eighteenth *Annual ACM-SIAM Symposium on Discrete Algorithms, ACM and SIAM,* (2007) 609-617.

23. F.Cheah and D.G.Corneil, On the structure of trapezoid graphs, *Discrete Applied Mathematics,* 66 (1996) 109-133.

24. E.Chen, L.Yang and H.Yuan, Improved algorithms for largest cardinality 2-interval pattern problem, *J. Comb. Optim.*, 13(3) (2007) 263-275.

25. D.G.Corneil, S.Olariu and L.Stewart, Asteroidal triple-free-graph, *SIAM J. Discrete Math.*, 10 (1997) 399-430.

26. D.G.Corneil and P.A.Kamula, Extensions of permutation and interval graphs, *Congress. Numer.,* 58 (1987) 267-275.

27. M.Crochemore, D.Hermelin, G.M.Landau, D.Rawitz and S.Vialette, Approximating the 2-interval pattern problem, *Theor. Comput. Sci.*, 395(2-3) (2008) 283-297.

28. I.Dagan, M.C.Golumbic and R.Y.Pinter, Trapezoid graphs and their coloring, *Discrete Appl. Math.*, 21 (1988) 35-46.

29. X.Deng, P.Hell and J.Huang, Linear time representation algorithms for proper circular-arc graphs and proper interval graphs, *SIAM J. Comput.*, 25 (1996) 390-403.

30. G.Dirac, On rigid circuit graphs, *Abh. Math. Sem. Univ. Hamburg*, 25 (1961) 71-76.

31. G.Durán, A.Gravano, R.M.McConnell, J.Spinrad and A.Tucker, Polynomial time recognition of unit circular-arc graphs, *J. Algorithms,* 58(1) (2006) 67-78.

32. G.Ehrlich, S.Even and R.E.Tarjan, Intersection graphs of curves in the plane, *Journal of Combinatorial Theory,* 21(1) (1976) 8-20.

33. F.V.Fomin, S.Gaspers, P.Golovach, K.Suchan, S.Szeider, E.J.van.Leeuwen, M.Vatshelle and Y.Villanger, $k$-gap interval graphs, LATIN 2012: *Theoretical Informatics, Lecture Notes in Computer Science*, 7256, (2012) 350-361.

34. D.R.Fulkerson and O.A.Gross, Incidence matrices and interval graphs, *Pacific J. Math.,* 15 (1965) 835-855.

35. M.R.Garey and D.S.Jhonson, *Computers and Interactibility: A Guide to the Theory of NP Completeness*, W. H. Freeman and Company, San Fransisco, (1979).

36. P.Galinier, M.Habib and C.Paul, Chordal graphs and their clique graphs, in: Springer-Verlag, editor, 21th Workshop on Graph-Theoretic Concepts in Computer Science, Aachen, *Lecture Notes in Computer Science,* 1017 (1995) 358-371.

37. F.Gavril, The intersection graphs of a path in a tree are exactly the chordal graphs, *J. Combinatorial Theory,* 16 (1974) 47-56.

38. A.Ghouilo-Houri, Characterisation des graphs non orientes dont on peut orienter les arretes de maniere a obtenir le graphe d'une relation d'ordre. *C. R. Acad. Sci.*, Paris, 254 (1962) 1370-1371.

39. R.Hayward, Weakly triangulated graphs, *J. Comb. Theory*, 39 (1985) 200- 208.

40. P.K.Ghosh and M.Pal, An optimal algorithm to solve 2-neighbourhood-covering problem on trapezoid graphs, *Advanced Modeling and Optimization*, 9(1) (2007) 15-36.

41. P.C.Gilmore and A.J.Hoffman, A characterization of comparability graphs and of interval graphs, *Canad. J. Math.*, 16 (1964) 539-548.

42. M.C.Gulumbic, *Algorithmic Graph Theory and Perfect Graphs*, Academic Press, New York (1980).

43. M.C.Gulumbic, C.L.Monma and W.T.Trotter, Tolerance graphs, *Discrete Applied Math.*, 9 (1984) 157-170.

44. M.C.Gulumbic and A.Trenk, Tolerance Graphs, Cambridge University Press, 2004.

45. J.R.Griggs and D.B.West, Extremal values of the interval number of a graph, *SIAM J. Algebra. Discr. Maths.*, 1(1) (1980) 1-14.

46. Hajös, G., Uber eine Art von Graphen, First posed the problem of characterizing interval graphs. *Intern. Math. Nachr.*, 11 (1957) , Problem 65.

47. D.Hermelin, J.Mestre and D.Rawitz, Optimization problems in dotted interval graphs, *Graph-Theoretic Concepts in Computer Science, Lecture Notes in Computer Science,* 7551 (2012) 46-56.

48. I.B.-A.Hartman, I.Newman and R.Ziv, On grid intersection graphs, *Discrete Math.*, 87 (1991) 41-52.

49. A.Hashimoto and J.Stevens, Wire routing by optimizing channel assignment within large apertures, in Proc., *8th IEEE Design Automation Workshop*, (1971) 155-169.

50. J.A.Horne and J.C.Smith, Dynamic programming algorithm for the conditional covering problem on path and extended star graph, *Networks*, 46(4) (2005) 177-185.

51. M.Hota, M.Pal and T.K.Pal, An efficient algorithm to generate all maximal independent sets on trapezoid graphs, *Intern. J. Computer Math.*, 70 (1999) 587-599.

52. M.Hota, M.Pal and T.K.Pal, An efficient algorithm for finding a maximum weight $k-$ independent set on trapezoid graphs, *Computational Optimization and Applications*, 18 (2001) 49-62.

53. M.Hota, M.Pal and T.K.Pal, Optimal sequential and parallel algorithms to compute all cut vertices on trapezoid graphs, *Computational Optimization and Applications*, 27 (2004) 95-113.

54. M.Hota, *Sequential and parallel algorithms on some problems of trapezoid graph*, Ph.D. Theis, Vidyasagar University, Midnapore, India, 2005.

55. S.F.Hwang and G.J.Chang, $k$ -neighbourhood-covering and independence problems for chordal graphs, *SIAM J. Discrete Math.*, 11(4) (1998) 633-643.

56. J.R.Jungck, O.Dick and A.G.Dick, Computer assisted sequencing, interval graphs and molecular evolution, *Biosystem*, 15 (1982) 259-273.

57. T.M.Kratzke and D.B.West, The total interval number of a graph II: Trees and complexity, *SIAM J. Discrete Math.*, 9(2) (1996) 339-348.

58. C.G.Lekkerkerker and J.C.Boland, Representation of a finite graph by a set of intervals on the real line, *Fund. Math.* 51 (1962) 45-64.

59. T.H.Ma and J.P.Spinrad, On the 2-chain subgraph cover and related problems, *J. Algorithms*, 17 (1994) 251-268.

60. T.A.McKee and F.R.McMorris, *Topics in Intersection Graph Theory, SIAM*, 1999.

61. S.Mandal and M.Pal, Maximum weight independent set of circular-arc graph and its application, *Journal of Applied Mathematics and Computing*, 22(3) (2006) 161-174.

62. S.Mandal and M.Pal, A sequential algorithm to solve next-to-shortest path problem on circular-arc graphs, *Journal of Physical Sciences*, 10 (2006) 201-217.

63. S.Mandal and M.Pal, An optimal algorithm to compute all hinge vertices on

circular-arc graphs, *Arab Journal of Mathematics and Mathematical Sciences*, 1(1) (2007) 16-27.

64. S.Mandal, A.Pal and M.Pal, An optimal algorithm to find centres and diameter of a circular-arc graph, *Advanced Modeling and Optimization*, 9(1) (2007) 155-170.

65. R.M.McConnell, Linear-time recognition of circular-arc graphs, *Algorithmica*, 37 (2) (2003) 93-147.

66. S.Mondal, M.Pal and T.K.Pal, An optimal algorithm for finding depth-first spanning tree on permutation graphs, *Korean J. of Computational and Applied Mathematics*, 6 (3) (1999) 493-500.

67. S.Mondal, M.Pal and T.K.Pal, An optimal algorithm for solving all-pairs shortest paths on trapezoid graphs, *International J. Computational Engineering Science*, 3(2) (2002) 103-116.

68. S.Mondal, M.Pal and T.K.Pal, An optimal algorithm to solve 2-neighbourhood -covering problem on interval graphs, *Int. J. Comput. Math.*, 79 (2002) 189-204.

69. S.Mondal, M.Pal and T.K.Pal, Optimal sequential and parallel algorithms to compute a Steiner tree on permutation graphs, *International J. Computer Mathematics*, 80(8) (2003) 937-943.

70. S.Mondal, M.Pal and T.K.Pal, An optimal algorithm to solve the all-pairs shortest paths problem on permutation graph, *J. Mathematical Modelling and Applications*, 2(1) (2003) 57-65.

71. S.Mondal, M.Pal and T.K.Pal, An optimal algorithm to solve the all-pairs shortest paths problem on permutation graph, *J. Mathematical Modelling and Applications*, 2(1) (2003) 57-65.

72. N.Nash and D.Gregg, An output sensitive algorithm for computing a maximum independent set of a circle graph, *Information Processing Letters*, 116(16) (2010) 630-634.

73. J.Fabri, *Automatic Storage Optimization*, UMI Press Ann Arbor, MI, 1982.

74. T.Ohtsuki, H.Mori, E.S.Khu, T.Kashiwabara and T.Fujisawa, One dimensional logic gate assignment and interval graph, *IEEE Trans. Circuits and Systems*, 26 (1979) 675-684.

75. J.Pach and G.T´oth, Recognizing string graphs is decidable, *Discrete and Computational Geometry,* 28 (2002) 593-606.

76. M.Pal and G.P.Bhattacharjee, An optimal parallel algorithm for computing all maximal cliques of an interval graph and its applications, *J. of Institution of Engineers (India)*, 76 (1995) 29-33.

77. M.Pal, *Some sequential and parallel algorithms on interval graphs*, Ph.D Thesis, Indian Institute of Technology, Kharagpur, India, 1995.

78. M.Pal and G.P.Bhattacharjee, Optimal sequential and parallel algorithms for computing the diameter and the centre of an interval graph, *Intern. J. Computer Maths.*, 59 (1995) 1-13.

79. M.Pal and G.P.Bhattacharjee, An improved algorithm for finding the maximum weight $k-$independent set on an interval graph, in Proc.: *5th National Seminar on Theoretical Computer Science,* Bombay, India, Aug. 1-4, (1995) 95-104.

80. M.Pal and G.P.Bhattacharjee, The parallel algorithms for determining edge-packing and efficient edge dominating sets in interval graphs, *Parallel Algorithms and Applications,* 7 (1995) 193-207.

81. M.Pal and G.P.Bhattacharjee, A sequential algorithm for finding a maximum weight $k$ -independent set on interval graphs, *Intern. J. Computer Maths.*, 60 (1996) 439-449.

82. M.Pal, An efficient parallel algorithm for computing a maximum-weight independent set of a permutation graph, in Proc.: *6th National Seminar on Theoretical Computer Science*, Banasthali Vidyapith, Rajasthan, India, Aug. 5-8 (1996) 276-285.

83. M.Pal and G.P.Bhattacharjee, An optimal parallel algorithm to color an interval graph, *Parallel Processing Letters*, 6(4) (1996) 439-449.

84. M.Pal and G.P.Bhattacharjee, A data structure on interval graphs and its applications, *J. Circuits, Systems, and Computer*, 7 (1997) 165-175.

85. M.Pal and G.P.Bhattacharjee, An optimal parallel algorithm for all-pairs shortest paths on unweighted interval graphs, *Nordic J. Computing*, 4 (1997) 342-356.

86. M.Pal, Efficient algorithms to compute all articulation points of a permutation graph, *Korean J. of Computational and Applied Mathematics,* 5 (1998) 141-152.

87. M.Pal, A parallel algorithm to generate all maximal independent sets on permutation graphs, *Intern. J. Computer Maths.*, 67 (1998) 261-274.

88. M.Pal, S.Mondal, D.Bera and T.K.Pal, An optimal parallel algorithm for computing cutvertices and blocks on interval graphs, *Intern. J. Computer Math.*, 75 (2000) 59-70.

89. R.Peeters, On coloring $j$ -unit spheres, Tech. Report FEW 512, Tilburg University, (1991).

90. A.Pnueli, S.Even and A.Lempel, Transitive orientation of graphs and identification of permutation graphs, *Canad. J. Math.*, 23 (1971) 160-175.

91. G.Ramalingam and C.Pandu Rangan, A unified approach to domination problem in interval graphs, *Information Processing Letters*, 27 (1988) 271-274.

92. A.Rana, A.Pal and M.Pal, The conditional covering problem on unweighted interval graphs, *Journal of Applied Mathematics and Informatics*, 28 (1-2) (2010) 1-11.

93. A.Rana, A.Pal and M.Pal, The 2-neighbourhood covering problem on permutation graphs, *Advanced Modelling and Optimization*, 13(3) (2011) 463-476.

94. A.Rana, A.Pal and M.Pal, The conditional covering problem on interval graphs with unequal costs, *Tamsui Oxford Journal of Information and Mathematical Sciences*, 27 (2) (2011) 183-195.

95. A.Rana, A.Pal and M.Pal, Efficient algorithms to solve k-domination problem on permutation graphs, Y.Wu (Ed.), *ICHCC 2011, CCIS 163*, Springer-Verlag, Berlin, Heidelberg, 327-334, 2011.

96. A.Rana, A.Pal and M.Pal, The conditional covering problem on trapezoid graphs, *ISRN Discrete Mathematics*, DOI: 10.5402/2011/213084. Volume 2011, Article ID 213084, 10 pages.

97. F.S.Roberts, On the boxicity and cubicity of a graph, In *Recent Progresses in Combinatorics*, Academic Press, 1969.

98. A.C.M.van Rooij and H.S.Wilf, The interchange graph of a finite graph, *Acta Mathematica Hungarica*, 16 (3–4) (1965) 263-269.

99. D.Rose, R.Tarjan and G.Lueker, Algorithmic aspects of vertex elimination on graphs, *SIAM J. of Computing*, 5 (1976) 266-283, 301-310.

100. A.Saha and M.Pal, Maximum weight $k$ -independent set problem on permutation graphs, *International J. of Computer Mathematics*, 80(12) (2003) 1477-1487.

101. A.Saha, M.Pal and T.K.Pal, An optimal parallel algorithm to find 3-tree spanner of interval graph, *International J. Computer Mathematics*, 82(3) (2005) 259-274.

102. A.Saha and M.Pal, An algorithm to find a minimum feedback vertex set of an interval graph, *Advanced Modeling and Optimization*, 7(1) (2005) 99-116.

103. A.Saha, M.Pal and T.K.Pal, An optimal parallel algorithm to find all-pairs shortest paths on circular-arc graphs, *J. Applied Mathematics and Computing,* 17 (1-2) (2005) 1-23.

104. A.Saha, M.Pal and T.K.Pal, An efficient PRAM algorithm for maximum weight independent set on permutation graphs, *Journal of Applied Mathematics and Computing,* 19 (1-2) (2005) 77-92.

105. A.Saha, M.Pal and T.K.Pal, Selection of programme slots of television channels for giving advertisement: A graph theoretic approach, *Information Sciences*, 177 (2007) 2480-2492.

106. E.R.Scheinerman and D.B.West, The interval number of a planar graph: Three intervals suffice, *J. Comb. Theory Ser. B*, 35(3) (1983) 224-239.

107. L.S.Skeide, *Recognizing weakly chordal graphs*, Ph.D. thesis, University of Bergen, Norway, (2002).

108. J.Spinrad, On comparability and permutation graphs, *SIAM J. Comput.*, 14 (1985) 658-670.

109. J.Spinrad, Recognition of circle graphs, *Journal of Algorithms*, 16(2) (1994) 264-282.

110. W.T.Trotter and F.Harary, On double and multiple interval graphs, *J. Graph Theory*, 3(3) (1979) 205-211.

111. A.Tiskin, Fast distance multiplication of unit-Monge matrices, *Proceedings of ACM-SIAM SODA* 2010, (2010) 1287–1296.

112. A.Tucker, Matrix characterrizations of circular-arc graphs, *Pacific J. Math.*, 38 (1971) 535-545.

113. A.Tucker, Structure theorems for some circular-arc graphs, *Discrete Mathematics*, 7 (1974) 167-195.

114. C.-W.Yu and G.-H.Chen, Generation of all maximal independent sets in permutation graphs, *Intern. J. Comput. Math.*, 47 (1993) 1-8.